\documentclass{tlp}
\usepackage{latexsym}
\usepackage{amssymb}
\usepackage{graphicx}
\usepackage{amsmath}
\usepackage{verbatim}
\usepackage{varioref}

\submitted{15 July 2004} \revised{1 August 2005} \accepted{13
January 2006}

\newtheorem{definition}{Definition} 
\newtheorem{example}{Example} 

\def\square{\hbox{\vrule\vbox{\hrule\phantom{o}\hrule}\vrule}}

\newcommand{\myboxi}{\,\hbox{\small $\square$}\,}

\newcommand{\Oz}{{\cal Oz}\,}
\newcommand{\toy}{\cal{TOY}}
\newcommand{\Toy}{{\cal TOY}\,}
\newcommand{\toyfd}{\cal{TOY(FD)}}
\newcommand{\Toyfd}{\cal{TOY(FD)}\,}

\newcommand{\TVar}{{\mathcal T\!V\!\!ar}}
\newcommand{\Cflp}{CFLP\,}
\newcommand{\cflpfd}{CFLP(\cal FD)}
\newcommand{\Cflpfd}{CFLP(\cal FD)\,}
\newcommand{\clpfd}{CLP(\cal FD)}
\newcommand{\Clpfd}{CLP(\cal FD)\,}

\newcommand{\restrict}{\!\restriction }

\newcommand{\Exp}{\mathit{Exp}}

\newcommand{\TSubst}{\mathit{TSubst}}
\newcommand{\Type}{\mathit{Type}}

\newcommand{\nat}{\mathbb N} %

\newcommand{\utup}{\overline{U}} %

\newcommand{\ex}{{\exists}}

\newcommand{\red}{\vdash\!\!\vdash} 

\newcommand{\tp}[1]{\overline{#1}}   

\newcommand{\tpp}[2]{\tp{#1}_{#2}}  

\newcommand{\expr}[1]{Exp_{\bot}(#1)}    

\newcommand{\pat}[1]{Pat_{\bot}(#1)} 

\newcommand{\sub}[1]{Sub_{\bot}(#1)}    

\newcommand{\nc}{\newcommand}

\nc{\val}[1]{Val_{\bot}(#1)}  

\nc{\tval}[1]{Val(#1)}        

\nc{\Var}{{\mathcal V\!ar}}  
\nc{\varx}{\mathcal{X}}    %
\nc{\vary}{\mathcal{Y}}    
\nc{\varz}{\mathcal{Z}}    %

\nc{\cdom}{\mathcal{FD}}

\nc{\sig}{\Sigma}

\nc{\ent}{\mathbb{Z}} \nc{\real}{\mathbb{R}}

\nc{\csig}{\Gamma}              
\nc{\pfun}[1]{PF^{#1}}          
\nc{\df}[1]{DF^{#1}}            
\nc{\uni}[1]{D_{#1}}            

\nc{\leqinfo}{\sqsubseteq} \nc{\lsinfo}{\sqsubset}
\nc{\geqinfo}{\sqsupseteq} \nc{\gtinfo}{\sqsupset}

\nc{\patom}{p\, \overline{t}_{n} \to!\, t\,}

\nc{\datom}{p\, \overline{e}_n \to!\, t\,}

\nc{\pcon}[1]{PCon_{\bot}(#1)}    
\nc{\ptcon}[1]{PCon(#1)}          
\nc{\pgcon}[1]{PGCon_{\bot}(#1)}  
\nc{\pgtcon}[1]{PGCon(#1)}        

\nc{\dcon}[1]{DCon_{\bot}(#1)}    
\nc{\dtcon}[1]{DCon(#1)}          
\nc{\dgcon}[1]{DGCon_{\bot}(#1)}  
\nc{\dgtcon}[1]{DGCon(#1)}        

\nc{\sol}[2]{Sol_{#1}(#2)}      

\nc{\true}{\lozenge} \nc{\false}{\blacklozenge}

\nc{\yields}{\to!\,}

\nc{\clp}[1]{CLP(#1)}    
\nc{\fp}[1]{FP(#1)}      
\nc{\cflp}[1]{CFLP(#1)}  

\nc{\ans}{\Pi \myboxi \theta}

\nc{\dem}[1]{dvar_{\cdom}(#1)}

\nc{\cproves}[1]{\vdash_{#1}} 

\nc{\ure}{\mathcal{U}} \nc{\var}{\mathcal{V}}
\nc{\implie}[1]{\models_{#1}}  
\nc{\prules}[1]{[#1]_{\bot}} 
\nc{\crwl}[1]{CRWL(#1)} 
\nc{\clnc}[1]{CLNC(#1)}  

\nc{\cdomd}{\mathcal{D}}

\begin{document}

\title[Constraint Functional Logic Programming over Finite Domains]
      {Constraint Functional Logic Programming over Finite Domains}

\author[~ A.J. Fern\'{a}ndez,T. Hortal\'{a}-Gonz\'{a}lez,F. S\'{a}enz-P\'{e}rez
and R. del Vado-V\'{\i}rseda] {ANTONIO J. FERN\'{A}NDEZ
\thanks {This author was partially
supported by Spanish MCyT under contracts
TIN2004-7943-C04-01 and TIN2005-08818-C04-01.} \\
      Dpto. de Lenguajes y Ciencias de la Computaci\'{o}n, \\
      Universidad de M\'{a}laga, Spain \\
      E-mail: afdez@lcc.uma.es
\and TERESA HORTAL\'{A}-GONZ\'{A}LEZ\footnotemark, \and FERNANDO
S\'{A}ENZ-P\'{E}REZ\addtocounter{footnote}{-1}
\thanks{The work of this author has been supported by the Spanish project PR
48/01-9901 funded by UCM.} and RAFAEL DEL VADO-V\'{I}RSEDA
\thanks{The work of this author has been partially supported by the Spanish National
Project MELODIAS (TIC2002-01167).}\\
      Dpto. de Sistemas Inform\'aticos y Programaci\'on \\
      Universidad Complutense de Madrid, Spain \\
      E-mails: {\{teresa,fernan,rdelvado\}@sip.ucm.es}
      }

\pagerange{\pageref{firstpage}--\pageref{lastpage}}
\setcounter{page}{1} \pubyear{2006}

\maketitle

\label{firstpage}

%
%

\begin{abstract}
In this paper, we present our proposal to Constraint Functional
Logic Programming over Finite Domains ($\Cflpfd$) with a lazy
functional logic programming language which seamlessly embodies
finite domain (${\cal FD}$) constraints. This proposal increases
the expressiveness and power of constraint logic programming over
finite domains ($\Clpfd$) by combining functional and relational
notation, curried expressions, higher-order functions, patterns,
partial applications, non-determinism, lazy evaluation, logical
variables, types, domain variables, constraint composition, and
finite domain constraints.

We describe the syntax of the language, its type discipline, and
its declarative and opera\-tional semantics. We also describe
$\toyfd$, an implementation for $\Cflpfd$, and a comparison of our
approach with respect to $\clpfd$ from a programming point of
view, showing the new features we introduce. And, finally, we show
a performance analysis which demonstrates that our implementation
is competitive with respect to existing $\clpfd$ systems and that
clearly outperforms the closer approach to $\Cflpfd$.
\end{abstract}


%
%

\section{Introduction}

Constraint logic programming ($CLP$)
\cite{jaffar+:clp-survey-jlp94} was born from a desire to have
both a better problem expression and performance than logic
programming ($LP$). Its success lies in that it combines the
declarativeness of $LP$  with the efficiency of the constraint
programming ($CP$) paradigm.  The essential component of the $CLP$
schema is that it can be parameterized by a computational domain
so that different domains determine different instances of the
schema. Constraint Programming over finite domains ($CP({\cal
FD})$) \cite{marriot+:constraints-mit98,henz+:fd-overview-apors00}
has emerged as one of the most exciting paradigms of programming
in recent decades.   There are several reasons for the interest
that $CP({\cal FD})$ has raised: (1) the strong theoretical
foundations
\cite{tsang:foundations-cs-93,apt:principles-constraint-programming-cambridge03,fruthwirth+:essentials-cp-springer2003}
that make $CP$ a sound programming paradigm; (2) $CP({\cal FD})$
is a heterogeneous field of research ranging from theoretical
topics in mathematical logic to practical applications in industry
(particularly, problems involving heterogeneous constraints and
combinatorial search) and (3) $CP$ is based on posing constraints,
which are basically true relations among domain variables. For
this last reason, the declarative languages seem to be more
appropriate than imperative languages to formulate ${\cal FD}$
constraint problems. In particular, one of the most successful
instances of $CP({\cal FD})$ is $\Clpfd$.

Another well-known instance of declarative programming ($DP$) is
functional programming ($FP$). The basic operations in functional
languages are defined using functions which are invoked using
function applications and put together using function composition.
$FP$ gives great flexibility, different from that provided by
$(C)LP$, to the programmer, because functions are first-class
citizens, that is, they can be used as any other object in the
language (i.e., results, arguments,  elements of data structures,
etc). Functional languages provide evident advantages such as
declarativeness, higher-order functions, polymorphism and lazy
evaluation, among others. To increase the performance, one may
think of integrating ${\cal FD}$ constraints into $FP$ (as already
done in $LP$). However, literature lacks proposals in this sense
and the reason seems to lie in the relational nature of ${\cal
FD}$ constraints, which do not fit well in $FP$. In spite of this
limitation, it seems clear that the integration of ${\cal FD}$
constraints into $FP$ is interesting not only for $FP$ but also
for discrete constraint programming, as the constraint community
may benefit from the multiple advantages of $FP$.

More recently, functional logic programming ($FLP$) emerges with
the aim to integrate the declarative advantages from both $FP$
and $LP$. $FLP$ gives rise to new features which
 cannot be found neither in $FP$ nor in $LP$
\cite{hanus:integrationfunc-logic-jlp97}. $FLP$
 has not the inherent limitations of $FP$ and thus it is an adequate
framework for the integration of functions and constraints. To our
best knowledge, the first  proposal for a constraint functional
logic programming scheme ($CFLP$) that attempts to combine
constraint logic programming and functional logic programming is
\cite{darlington+:perspective-integration-funt-logc-language-fgcs92}.
The idea behind this approach can be roughly described by the
equation $\cflp{\cdomd} = \clp{\fp{\cdomd}}$, intended to mean
that a $\Cflp$ language over the constraint domain $\cdomd$ is
viewed as a $CLP$ language over an extended constraint domain
$\fp{\cdomd}$ whose constra\-ints include equations between
expressions involving user defined functions, which will be solved
via narrowing. Further, the $\Cflp$ scheme proposed by F.J.
L\'{o}pez-Fraguas in \cite{fraguas:general-schema-cflpX-alp92}
tried to provide a declarative semantics such that $\clp{\cdomd}$
programs could be formally understood as a particular case of
$\cflp{\cdomd}$ programs. In the classical approach to $CLP$
semantics,  a constraint domain is viewed as a first-order
structure $\cdomd$, and constraints are viewed as first-order
formulas that can be interpreted in $\cdomd$. In
\cite{fraguas:general-schema-cflpX-alp92}, $\cflp{\cdomd}$
programs were built as sets of constrained rewriting rules. In
order to support a lazy semantics for the user defined functions,
constraint domains $\cdomd$ were formalized as continuous
structures, with a Scott domain \cite{gunter+:semantic-domains90}
as a carrier, and a continuous interpretation of function and
predicate symbols. The resulting semantics had many pleasant
properties, but also some limitations. In particular, defined
functions had to be first-order and deterministic, and the use of
patterns in function definitions had to be simulated by means of
special constraints.

In a recent work
\cite{fraguas+:constraint-functional-logic-progr-rev-wrla04}, a new
generic scheme $\cflp{\cdomd}$ has been proposed, intended as a
logical and semantic framework for lazy Constraint Functional Logic
Programming over a parametrically given constraint domain $\cdomd$,
which provides a clean and rigorous declarative semantics for
$\cflp{\cdomd}$ languages as in the $\clp{\cdomd}$ scheme,
overcoming the limi\-tations of the older $\cflp{\cdomd}$ scheme
\cite{fraguas:general-schema-cflpX-alp92}. $\cflp{\cdomd}$ programs
are presented as sets of constrained rewriting rules that define the
behavior of possibly higher-order and/or non-deterministic lazy
functions over $\cdomd$. The main novelties in
\cite{fraguas+:constraint-functional-logic-progr-rev-wrla04} were a
new formalization of constraint domains for $\Cflp$, a new notion of
interpretation for $\cflp{\cdomd}$ programs, and a new Constraint
Rewriting Logic $\crwl{\cdomd}$ parameterized by a constraint
domain, which provides a logi\-cal characterization of program
semantics. Further,
\cite{fraguas+:lazy-narrowing-calculus-for-declarative-cp-ppdp04}
has formalized an operational semantics for the new generic scheme
$\cflp{\cdomd}$ proposed in
\cite{fraguas+:constraint-functional-logic-progr-rev-wrla04}. This
work presented a constrained lazy narrowing calculus $\clnc{\cdomd}$
for solving goals for $\cflp{\cdomd}$ programs, which was proved
sound and strongly complete with respect to $\crwl{\cdomd}$'s
semantics. These properties qualified $\clnc{\cdomd}$ as a
convenient computation mechanism for declarative constraint
programming languages. More recently, \cite{vado+:frocos05}
presented an optimization of the $\clnc{\cdomd}$ calculus by means
of definitional trees \cite{antoy:definitional-trees-alp-92} to
efficiently control the computation. This new constrained demanded
narrowing calculus $CDNC(\cdomd)$ preserves the soundness and
completeness properties of $\clnc{\cdomd}$ and maintains the good
properties shown for needed and demand-driven narrowing strategies
\cite{DBLP:conf/plilp/LoogenFR93,hanus:+needed,vado+:ppdp03}. These
properties adequately render $CDNC(\cdomd)$ as a concrete
specification for the implementation of the computational behavior
in existing $\cflp{\cdomd}$ systems such as $\mathcal{TOY}$
\cite{toy-manual97} and $Curry$ \cite{hanus:curry99}.

The main contributions of the paper are listed below:
\begin{itemize}
\item The paper describes the theoretical foundations for the
$\Cflpfd$ language, i.e., a concrete instance of the general
scheme $\cflp\cdomd$ presented in
\cite{fraguas+:constraint-functional-logic-progr-rev-wrla04,fraguas+:lazy-narrowing-calculus-for-declarative-cp-ppdp04}.
First, this instance includes an explicit treatment of a type
system for constraints as well as for programs, goals and answers.
Second, it also presents a new formalization of the constraint
finite domain $\cdom$ for $\Cflp$ that includes a succinct
declarative semantics (similarly as done for $CLP$) for an
enough-expressive primitive constraints set. Finally, it provides
the formal specification of a finite domain constraint solver
defined over those primitive constraints that constitutes  the
theoretical basis for the implementation of correct propagation
solvers.

\item The paper presents an operational semantics for finite
domain constraint solving on declarative languages using a new
constraint lazy narrowing calculus $CLNC(\cdom$), consisting of a
natural and novel combination of lazy evaluation and ${\cal FD}$
constraint solving that does not exist, to our knowledge, in any
declarative constraint solver (see
\cite{fraguas+:lazy-narrowing-calculus-for-declarative-cp-ppdp04}
and Section \ref{sect: related work}). This operational semantics is
defined with respect to a constraint rewriting logic over a ${\cal
FD}$ setting that makes it very different from the operational
behavior of $\Clpfd$ languages.

\item The paper presents $\Toyfd$ from a programming point of view,
which is the first $\Cflpfd$ system that
provides a wide set of ${\cal FD}$ constraints comparable to
existing $\Clpfd$ systems and which is competitive with them, as
shown with performance results. Also, the advantages of combining
${\cal FD}$ constraints into $FLP$ are highlighted via examples.
Our system is therefore a contribution for increasing the
expressiveness and efficiency of $FLP$ by using ${\cal FD}$
constraints and a state-of-the-art propagation solver.
\end{itemize}

The structure of the paper is as follows. Section~\ref{sect:
CFLPFD programs} presents our $\Cflpfd$ language by describing its
type discipline, patterns and expressions, finite domains, and
constraint solvers. In Section~\ref{sect: object}, we provide a
constraint lazy narrowing calculus over ${\cal FD}$ domains
($CLNC(\cdom)$) along with the notions of well-typed programs,
admissible goals, and correct answers. Next, Section
\ref{sect:TOYFD} describes our implementation of $\cflpfd$:
$\toyfd$, highlighting the advantages obtained from embodying
constraints into a functional logic language with respect to
$\Clpfd$, and comparing the performance of our $\Cflpfd$ system
with other declarative constraint systems. Section~\ref{sect:
related work} discusses related work and, finally, Section
\ref{Conclusions and Future Work} summarizes some conclusions and
points out future work.

%
%
%

\section{The $\Cflpfd$ Language}\label{sect: CFLPFD programs}

We propose a constraint functional logic programming language over
finite domains which is a pure declarative language, typed, lazy,
and higher-order, and that provides support for constraint solving
over finite domains. $\Cflpfd$ aims to the integration of the best
features of existing functional and logic languages into ${\cal FD}$
constraint solving.

In this section, we present the basis of our $\Cflpfd$ language
about syntax, type discipline, and declarative semantics. We also
formalize the notion of a constraint finite domain and specify the
expected behavior that a ${\cal FD}$ constraint solver attached to
our $\Cflpfd$ language must hold.

\subsection{Polymorphic Signatures} \label{types syntax}

We assume a countable set  $\TVar$ of \emph{type variables}
$\alpha,\ \beta,\ \ldots$  and a countable ranked alphabet $TC =
\bigcup_{n \in {\mathbb N}}TC^n$ of \emph{type constructors} $C
\in TC^n$. Types $\tau  \in \Type$ have the syntax  $\tau ::=
\alpha~~\mid C~\tau_{1}\ldots\tau_{n}~~\mid \tau \to \tau' \mid
(\tau_1, \dots, \tau_n)$.

By convention, $C ~\overline{\tau}_{n}$ abbreviates $C~\tau_{1}
\ldots\tau_{n}$, ``$\to$'' associates to the  right,
$\overline{\tau}_{n}\to \tau$ abbreviates $\tau_{1} \to \cdots \to
\tau_{n} \to \tau$, and the set of type variables occurring in
$\tau$ is written $tvar(\tau)$. A type $\tau$  is called {\em
monomorphic} iff $tvar(\tau) = \emptyset$, and {\em polymorphic}
otherwise. $(\tau_1, \dots, \tau_n)$ is a type intended to denote
$n$-tuples. A type without any occurrence of ``$\to$'' is called a
{\em datatype}. {\em Datatype} definitions declare new (possibly
polymorphic) {\em constructed types} and determine a set of {\em
data constructors} for each type. Our $\Cflpfd$ language provides
predefined types such as $[A]$ (the type of polymorphic lists, for
which Prolog notation is used), $bool$ (with constants $true$ and
$false$), $int$ for integer numbers, $char$ (with constants
$a$, $b$, \ldots) or $success$ (with constant $\top$).

\noindent A {\em polymorphic signature} over $TC$ is a triple
$\Sigma = \langle TC,\ DC,\ FS \rangle$, where $DC = \bigcup_{n
\in {\mathbb N}} DC^n$ and $FS = \bigcup_{n \in {\mathbb N}} FS^n$
are ranked sets of {\em data constructors} and {\em evaluable
function symbols}. Evaluable functions can be further classified
into domain dependent {\em primitive functions} $PF^n$ $\subseteq$
$FS^n$ and user {\em defined functions} $DF^n$ $=$ $FS^n$
$\backslash$ $PF^n$ for each $n \in \nat$.

Each $n$-ary $c \in DC^n$ comes with a principal type declaration
$c~::~\overline{\tau}_n \to C~\overline{\alpha}_k$, where $n,
k\geq 0, \alpha_{1},\ldots,\alpha_{k}$ are pairwise different,
$\tau_{i}$ are datatypes, and $tvar(\tau_{i}$) $\subseteq$
\{$\alpha_{1}$,\ldots, $\alpha_{k}$\} for all $1 \le i \le n$.
Also, every $n$-ary $f \in FS^n$ comes with a principal type
declaration $f~::~ \overline{\tau}_n \to \tau$, where $\tau_{i}$,
$\tau$  are arbitrary types. For any polymorphic signature
$\Sigma$, we write $\Sigma_{\bot}$ for the result of extending
$DC^0$ in $\Sigma$ with a special data constructor $\bot$,
intended to represent an undefined value that belongs to every
type. We also assume that $DC^0$ includes the three constants
mentioned above $true, ~ false ~::~bool$, and $\top~::~success$,
which are useful for representing the results returned by various
primitive functions. As notational conventions, in the rest of the
paper, we use $c, d \in DC$, $f, g \in FS$ and $h \in DC \cup FS$,
and we define the arity of $h \in DC^{n} \cup FS^{n}$ as $ar(h) =
n$.

\subsection{Expressions, Patterns and Substitutions}

In the sequel, we always assume a given polymorphic signature
$\Sigma$, often not made explicit in the notation. We introduce
the syntax of applicative expressions and patterns, which is
needed for understanding the construction of constraint finite
domains and constraint finite domain solvers. We assume a
countably infinite set $\Var$ of {\em (data) variables} $X, Y,
\ldots$ and the integer set $\mathbb{Z}$ of primitive elements
$0,$ $1,$ $-1,$ $2,$ $-2,$ $\ldots,$ mutually disjoint and
disjoint from $\TVar$ and $\Sigma$. Primitive elements are
intended to represent the finite domain specific set of integer
values.

An {\em expression} $e \in \expr{\mathbb{Z}}$ has the syntax $e
::= \, \perp \, \mid u \, \mid \, X\, \mid h \, \mid \, (e\, e')
\, \mid (e_1,\ \dots,\ e_n)$ where $u \in \mathbb{Z}$, $X \in
\Var$, $h$ $\in$ $DC$ $\cup$ $FS$, $(e\, e')$ stands for the {\em
application} operation which applies the function denoted by $e$
to the argument denoted by $e'$ and $(e_1, \dots, e_n)$ represents
{\em tuples} with $n$ components. The set of data variables
occurring in $e$ is written $var(e)$. Moreover, $e$ is called {\em
linear} iff every $X$ $\in$ $var(e)$ has a single occurrence in
$e$, {\em ground} iff $var(e)$ $=$ $\emptyset$ and {\em total} iff
is an expression with no occurrences of $\bot$. {\em Partial
patterns} $t \in \pat{\mathbb{Z}} \subset \expr{\mathbb{Z}}$ are
built as $t ::= \, \bot \, \mid \, u \, \mid \, X \, \mid c\,
t_{1} \ldots t_{m}\, \mid \, f\,t_{1} \ldots t_{n}\,$ where $u \in
\mathbb{Z}$, $X \in \Var$, $c\in DC$ with $m\leq ar(c)$, and $f\in
FS$ with $n<ar(f)$. Notice that partial applications (i.e.,
application to less arguments than indicated by the arity) of $c$
and $f$ are allowed as patterns, which are then called {\em
higher-order patterns}
\cite{gonzalez+:polymorphic-types-FLP-flops99}, because they have
a functional type. We define the {\em information ordering}
$\leqinfo$ as the least partial ordering over $\pat{\mathbb{Z}}$
satisfying the following properties: $\bot \leqinfo t$ for all $ t
\in \pat{\mathbb{Z}}$ and $h\ \overline{t_m} \leqinfo h\
\overline{t'_m}$ whenever these two expressions are patterns and
$t_i \leqinfo t'_i$ for all $i \in \{1,\ldots,m\}$.

As usual, we define {\em (data) substitutions} $\sigma
\in\sub{\mathbb{Z}}$ as mappings $\sigma : \Var \to
\pat{\mathbb{Z}}$ extended to $\sigma : \expr{\mathbb{Z}} \to
\expr{\mathbb{Z}}$ in the natural way. By convention, we write
$\varepsilon$ for the identity substitution, $e \sigma$ instead of
$\sigma(e)$ and $\sigma\theta$ for the composition of $\sigma$ and
$\theta$. We define the domain $dom(\sigma)$ of a substitution
$\sigma$ in the usual way. A substitution $\sigma$ such that
$\sigma\sigma$ $=$ $\sigma$ is called {\em idempotent}. For any
set of variables $\varx$ $\subseteq$ $\Var$ we define the {\em
restriction} $\sigma$ $\restrict$ $\varx$ as the substitution
$\sigma'$ such that $dom(\sigma')$ $=$ $\varx$ and $\sigma'(X)$
$=$ $\sigma(X)$ for all $X$ $\in$ $\varx$. We use the notation
$\sigma$ $=_{\varx}$ $\theta$ to indicate that $\sigma$
$\restrict$ $\varx$ $=$ $\theta$ $\restrict$ $\varx$, and we
abbreviate $\sigma$ $=_{\Var \setminus \varx}$ $\theta$ as
$\sigma$ $=_{\setminus \varx}$ $\theta$. {\em Type substitutions}
can be defined similarly, as mappings $\sigma_{t} : \TVar\,
\rightarrow \, \Type$ with a unique extension $\hat{\sigma_{t}} :
\Type \rightarrow \Type$, denoted also as $\sigma_{t}$. The set of
all type substitutions is denoted as $\TSubst$. Most of the
concepts and notations for data substitutions (such as domain,
composition, etc.) make sense also for type substitutions, and we
will freely use them when needed.

\subsection{Well-typed Expressions}
Inspired by Milner's type system \cite{damas+:types} we now
introduce the notion of well-typed expression. We define a
\emph{type environment} as any set $T$ of type assumptions $X ::
\tau$ for data variables s.t. $T$ does not include two different
assumptions for the same variable. The domain $dom(T)$ of a type
environment is the set of all the data variables that occur in
$T$. For any variable $X \in dom(T)$, the unique type $\tau$ s.t.
$(X :: \tau) \in T$ is denoted as $T(X)$. The notation
$(h~::~\tau) \in_{var} \Sigma$ is used to indicate that $\Sigma$
includes the type declaration $h~::~\tau$ up to a renaming of type
variables. {\em Type judgements} ${(\Sigma, T) \vdash_{WT}}\ e ::
\tau$ with $e \in Exp_{\perp}(\mathbb{Z})$ are derived by means of
the following {\em type inference} rules: \\\\
\noindent
\begin{tabular}{l}
 ${(\Sigma, T) \vdash_{WT}\ } u :: int$, for every
$u \in \mathbb{Z}$.\\
 ${(\Sigma, T) \vdash_{WT}\ } X :: \tau$, if $T(X) = \tau$.\\
 ${(\Sigma, T) \vdash_{WT}\ } h :: \tau\sigma_{t}$, if
$(h :: \tau) \in_{var} \Sigma_{\bot}, ~\sigma_{t} \in \TSubst$.\\
 ${(\Sigma, T) \vdash_{ WT}\ } (e ~e') :: \tau$, if
${(\Sigma,T) \vdash_{WT}\ } e :: (\tau' \to \tau)$,
${(\Sigma, T) \vdash_{ WT}\ } e':: \tau'$, \\
    \hspace*{3.5cm}for some $\tau' \in \Type$.\\
${(\Sigma, T) \vdash_{ WT}\ } (e_1, \dots,\! e_n)\!
::\! (\tau_1,\dots,\!\tau_n)$, if $\forall i \in
\{1,\ldots,\!n\}\!:\!{(\Sigma, T) \vdash_{WT}\ }\! e_i\!::\!
\tau_{i}$.
\end{tabular} \\

An expression $e \in \Exp_\perp$ is called {\em well-typed} iff
there exist some {\em type environment} $T$ and some type $\tau$,
such that the {\em type judgement} ${(\Sigma, T) \vdash_{ WT}}\ e ::
\tau$ can be derived. Expressions that admit more than one type
are called {\em polymorphic}. A well-typed expression always
admits a so-called {\em principal type} (PT) that is more general
than any other. A pattern whose PT determines the PTs of its
subpatterns is called {\em transparent}.

\subsection{The Constraint Finite Domain $\cdom$} \label{domains}

Adopting the general approach of
\cite{fraguas+:constraint-functional-logic-progr-rev-wrla04,fraguas+:lazy-narrowing-calculus-for-declarative-cp-ppdp04},
a {\em constraint finite domain} $\cdom$ can be formalized as a
structure with carrier set $\uni{\mathbb{Z}}$, consisting of
ground patterns built from the symbols in a polymorphic signature
$\sig$ and the set of primitive elements $\mathbb{Z}$. Symbols in
$\Sigma$ are intended to represent data constructors (e.g., the
list constructor), domain specific primitive functions (e.g.,
addition and multiplication over $\ent$) satisfying {\em
monotonicity}, {\em antimonotonicity} and {\em radicality}
properties (see
\cite{fraguas+:constraint-functional-logic-progr-rev-wrla04} for
details), and user defined functions. Requiring primi\-tives to be
{\em radical}, which just means that for given arguments, they are
expected to return a total result, unless the arguments bear too
few information for returning any result different from $\bot$. As
we will see in the next subsection, it is also possible to
instantiate this notion of constraint finite domain by adding a
new formal specification of a constraint finite domain solver
$Solve^{\cdom}$ as the theoretical basis of our operational
semantics and implementation.

Assuming then a constraint finite domain $\cdom$, we define first
the syntax and semantics of constraints over $\cdom$ used in this
work. In contrast to $CLP(\cdom)$, our constraints can include now
occurrences of user defined functions and can return any value of
the $\Type$ set.
\begin{itemize}
\item {\em Primitive constraints} have the syntactic form
$\patom$, being $p \in PF^{n}$ a primitive function symbol and
$t_1,\ldots,t_{n},t$ $\in$ $\pat{\mathbb{Z}}$ with $t$ total.
\item {\em Constraints} have the syntactic form $\datom$, with $p
\in PF^{n}$, $e_1,\ldots,e_{n} \in \expr{\mathbb{Z}}$ and $t \in
\pat{\mathbb{Z}}$ total.
\end{itemize}
In the sequel, we  use the notation $PCon(\cdom)$ for the set of
all the primitive constraints $\pi$ over $\cdom$. We reserve the
capital letter $\Pi$ for sets of primitive constraints, often
interpreted as conjunctions. The semantics of primitive
constraints depends on the notion of {\em valuation}
$\tval{\cdom}$ over $\cdom$, defined as the set of substitutions
of ground patterns for variables. The set of {\em solutions} of
$\pi$ $\in$ $PCon(\cdom)$ is a subset $\sol{\cdom}{\pi} \subseteq
\tval{\cdom}$ that satisfy $\pi$ in $\cdom$ in the sense of
\cite{fraguas+:constraint-functional-logic-progr-rev-wrla04}.
Analogously, the set of solutions of $\Pi \subseteq PCon(\cdom)$
is defined as $\sol{\cdom}{\Pi}$ $=$ $\bigcap_{\pi \in~ \Pi}$
$\sol{\cdom}{\pi}$. Moreover, we define the set of solutions of a
pair $\Pi$ $\myboxi$ $\sigma$ with $\sigma$ $\in$
$Sub_{\bot}(\ent)$ as $\sol{\cdom}{\Pi~\myboxi~\sigma}$ $=$
$\sol{\cdom}{\Pi}$ $\cap$ $Sol(\sigma)$, where $Sol(\sigma)$ is
the set of valuations $\eta$ such that $X\eta$ $\equiv$ $t\eta$
for each $X \mapsto t \in \sigma$.

The next definition presents a possible specific polymorphic
signature with finite domain constraints constituted by a minimum
set of primitive function symbols with their corresponding
declarative semantics. By means of this signature, our $\Cflpfd$
language allows the management of the usual finite domain
constraints defined over $\mathbb{Z}$ in $\Clpfd$ and also
equality and disequality constraints defined over
$Pat_{\bot}(\mathbb{Z})$ in a similar way as done in
\cite{gonzalez+:polymorphic-types-FLP-flops99}.\\

\begin{definition} \label{semantica}

Consider the following usual primitive operators and relations
defined over $\mathbb{Z}$:\\

$\otimes^{\mathbb{Z}}~::~  int ~\rightarrow~ int ~\rightarrow~
int$, ~ where $\otimes ~\in~ \{+,-,*,/\}$

$\asymp^{\mathbb{Z}}~::~ int ~\rightarrow~  int ~\rightarrow~
bool$, ~ where  $\asymp ~\in~ \{=,\neq,<,\leq,>,\geq\}$.

\begin{table}[htbp]
\caption{Primitive Function Symbols in $\cdom$ and their
Declarative Interpretation }\label{set of primitive constraints}
\begin{center}
\begin{small}
\begin{tabular}{ll} \hline\hline
{\bf Strict Equality}      & $seq:: ~ \alpha \to \alpha \to  bool$\\
{\bf(on patterns)}         & $seq^{\cdom} : \uni{\mathbb{Z}} \times \uni{\mathbb{Z}} \to  \{true, false,\bot\}$ \\
                           & $seq^{\cdom}  ~ t ~~ t  \to ~true$, $\forall t \in \uni{\mathbb{Z}}$ total   \\
                           & $seq^{\cdom}  ~ t_{1} ~ t_{2}  \to~ false, \forall t_1, t_2 \in \uni{\mathbb{Z}}$.
               $t_1, t_2$ have no common upper \\
               &\hspace*{3.2cm}bound  w.r.t. the information ordering $\sqsubseteq$\\
                           & $seq^{\cdom}  ~ t_{1} ~ t_{2}  \to~ \bot$, ~ otherwise   \\\hline

{\bf Less or Equal}        & $leq:: ~ int \to ~int ~ \to bool$\\
{\bf (on integers)}        & $leq^{\cdom} : \uni{\mathbb{Z}} \times \uni{\mathbb{Z}} \to \{true, false,\bot\}$ \\
                           & $leq^{\cdom}  ~ u_{1} ~ u_{2} \to~ true$,~  if $u_{1},u_{2} \in \mathbb{Z}$ and $u_{1} \leq^{\mathbb{Z}} u_{2}$  \\
                           & $leq^{\cdom}  ~ u_{1} ~ u_{2} \to~ false$, ~ if $u_{1},u_{2} \in \mathbb{Z}$ and $u_{1} >^{\mathbb{Z}} u_{2}$ \\
                           & $leq^{\cdom}  ~ u_{1} ~ u_{2}  \to~ \bot$, ~otherwise
                           \\\hline

{\bf Operators}            & $\otimes:: ~ int \to ~ int ~  \to int$\\
{\bf (on integers)}        & $\otimes^{\cdom} : \uni{\mathbb{Z}} \times \uni{\mathbb{Z}} \to \uni{\mathbb{Z}}$  \\
                           & $\otimes^{\cdom}  ~ u_{1} ~ u_{2} \to~ u_{1} \otimes^{\mathbb{Z}} u_{2}$, ~ if $u_{1},u_{2} \in \mathbb{Z}$ \\
                           & $\otimes^{\cdom}  ~ u_{1} ~  u_{2} \to~ \bot$,
                           ~otherwise\\\hline

{\bf  Finite Domains}      & $domain:: ~ int \to ~ [int] ~ \to bool$\\
                           & $domain^{\cdom} : \uni{\mathbb{Z}} \times \uni{\mathbb{Z}} \to \{true, false,\bot\}$  \\
                           & $domain^{\cdom}  ~ u ~ [u_{1}, \ldots,u_{n}]  \to ~ true$, \\
                           &\hspace*{0.5cm} if $\forall i \in \{1,\ldots,n\!-\!1 \} . u_{i} \leq^{\mathbb{Z}} u_{i+1}$ and $\exists i \in \{1,\ldots,n \}. u =^{\mathbb{Z}} u_{i}$ \\
                           & $domain^{\cdom}  ~ u ~ [u_{1}, \ldots,u_{n}]  \to ~false$, \\
                           & \hspace*{0.5cm} if $\exists i \in \{1,\ldots,n\!-\!1 \} . u_{i} >^{\mathbb{Z}} u_{i+1}$ or $\forall i \in \{1,\ldots,n \}. u \neq^{\mathbb{Z}} u_{i}$ \\
                           & $domain^{\cdom}  ~ u ~ [u_{1}, \ldots,u_{n}] \to \bot$, otherwise
                           \\\hline

{\bf  Variable Labeling}   & $indomain:: ~ int \to ~ success$\\
                           & $indomain^{\cdom} : \uni{\mathbb{Z}} \to \{\top, \bot\}$ \\
                           & $indomain^{\cdom}  ~  u  ~ \to~ \top$, ~ if $u \in \mathbb{Z}$ \\
                           & $indomain^{\cdom}  ~ u ~ \to~ \bot$, ~otherwise\\
\hline\hline
\end{tabular}
\end{small}
\end{center}
\end{table}

Table~\ref{set of primitive constraints} shows the set of
primitive functions $p$ $\in$ $PF^n$ with their corresponding type
declarations and their declarative interpretation $p^{\cdom}$
$\subseteq$ $D_{\ent}^{n}$ $\times$ $D_{\ent}$ considered in our
constraint finite domain $\cdom$ (we use the notation
$p^{\cdom}\tpp{t}{n}$ $\to$ $t$ to indicate that $(\tpp{t}{n},t)$
$\in$ $p^{\cdom}$). We note that all our primitive functions
satisfy the aforementioned properties.
 \hfill $\blacklozenge$

\end{definition}

The function $indomain$ is the basis for a {\em labeling} ({\em
enumeration} or {\em search}) strategy, which is crucial in
constraint solving efficiency. {\em labeling} is applied when no
more constraint propagation is possible, and its basic step consists
of selecting a variable $X$ with a non-empty, non-singleton domain,
selecting a value $u$ of this domain, and assigning $u$ to $X$. We
note that in our framework, various labeling strategies (variable
and value selection) have all the same declarative semantics, but
they may differ in their operational behavior and therefore in
efficiency as it happens in the $\Clpfd$ setting (more details can
be found in Section~\ref{labelings}). In the rest of the
paper, when opportune, we use the following notations:

\begin{itemize}
\item $t == s$ abbreviates $seq\ t\  s \to!~true$ and $t
~\backslash= s$ abbreviates $seq\ t\ s \to!~false$ (the notations
$=$ and $\neq$ can be understood as a particular case of the
notations $==$ and $\backslash=$ when these are applied to
integers and/or integer variables).
\item $a \leq b$ abbreviates $leq\ a\ b \to!~true$ (and
analogously for the other comparison primitives $<$, $>$ and
$\geq$).
\item $a \otimes b \asymp c$ abbreviates $a \otimes b \to!~r, r
\asymp c$.
\item $u \in  D$ abbreviates $domain\ u\ D \to!~true$ and
$u_1,\ldots,u_n \in D$ abbreviates $u_1 \in  D$ $\wedge$ $\ldots$
$\wedge$ $u_n \in D$. Analogously, $u \notin  D$ abbreviates
$domain\ u\ D \to!~false$ and $u_1,\ldots,u_n \notin D$
abbreviates $u_1 \notin  D$ $\wedge$ $\ldots$ $\wedge$ $u_n \notin
D$.
\item $domain$ $[u_1,\ldots,u_n]$ $a$ $b$ with $a,b$ $\in$ $\ent$
$(a \leq b)$ abbreviates $u_1,\ldots,u_n$ $\in$ $[a$ $..$ $b]$,
where $[a$ $..$ $b]$ represents the
integer list $[a,a+1,\ldots,b-1,b]$
that intuitively represents the integer interval [a,b].
\item $labeling$ $L$ $[u_{1},\ldots,u_{n}]$ abbreviates and
extends $indomain$ $u_{1}$ $\to!$ $\top$ $\land$ $\ldots$ $\land$
$indomain$ $u_{n}$ $\to!$ $\top$ with a list $L$ of predefined
constants that allow to specify different labeling strategies.
\end{itemize}
Using these notations, a {\em primitive constraint store} $S
\subseteq PCon(\cdom)$ can be expressed as a finite conjunction of
primitive constraints of the form $t == s$, $t ~\backslash= s$, $u
\in D$, $u \notin D$, $a \otimes b \asymp c$,
$domain~[u_1,\ldots,u_n]~a~b$ and/or $labeling~L~[u_1,\ldots,u_n]$
where $t, s$ are total patterns, $u_i,u, a, b, c~\in~\mathbb{Z}~\cup~\Var$,
and $L,D$ are total patterns representing a variable or a list.

\subsection{Constraint Solvers over $\cdom$} \label{solvers}

The design of a suitable operational semantics over finite domains
for goal solving in $\cflpfd$ combines constrained lazy narrowing
with constraint solving over a given constraint finite domain
$\cdom$. The central notion of lazy narrowing can be found in the
literature, e.g.,
\cite{middeldorp+:deterministic-lazy-narrowing-jsc98,DBLP:journals/jflp/MiddeldorpSH02}.
In this subsection, we describe the expected behavior of a
constraint solver over the finite constraint domain $\cdom$ w.r.t.
the semantics given in the previous subsection, as the basis of
our goal solving mechanism.

\begin{definition} \label{defsol}
A {\em constraint solver} over a given constraint domain $\cdom$ is
a function named $Solve^{\cdom}$ expecting as parameters a finite
primitive constraint store $S \subseteq \ptcon{\cdom}$ in the sense
of Definition \ref{semantica} and a finite set of variables $\chi
\subseteq \Var$ called the set of {\em protected variables}. The
constraint solver is expected to return a finite disjunction of $k$
alternatives: $Solve^{\cdom}(S,\chi)$ = $\bigvee_{i = 1}^{k}(S_{i}
\myboxi \sigma_{i})$, where each $S_{i} \subseteq \ptcon{\cdom}$ and
each $\sigma_{i} \in \sub{\mathbb{Z}}$ is a total idempotent
substitution satisfying the following requirements: no alternative
can bind protected variables, for each alternative either all the
protected variables disappear or some protected variable becomes
{\em demanded} (i.e., no solution can bind these variables to an
undefined value), no solution is lost by the constraint solver, and
the solution space associated to each alternative is included in one
of the input constraint stores (i.e., $Sol_{\cdom}(S) = \bigcup_{i =
1}^{k} Sol_{\cdom}(S_{i} \myboxi \sigma_{i})$). In the case $k = 0$,
the disjunction is understood as failure and $Sol_{\cdom}(S) =
\emptyset$ (that means failure detection).\hfill $\blacklozenge$
\end{definition}

\cite{fraguas+:lazy-narrowing-calculus-for-declarative-cp-ppdp04}
describes a constraint solver defined on the domain
$\mathcal{H}_{seq}$ in which the constraints considered are just
those for the strict (dis)equality on pure patterns (i.e. those
patterns constructed  over an empty set of primitive elements). Now,
in this paper, we extend this solver to the constraint domain
$\cdom$ in which we consider $\mathbb{Z}$ as the set of primitive
elements. We follow this approach and assume that the solver
$Solve^{\cdom}$ will behave as follows: $Solve^{\cdom}(S,\chi) =
\bigvee_{i=1}^{k} (S_{i} \myboxi \sigma_{i})$ iff $S \myboxi \,
\varepsilon \red_{\chi}^{*} \bigvee_{i=1}^{k}(S_{i} \myboxi
\sigma_{i}) \not\red_{\chi}$, where the relation $\red_{\chi}$
expresses a solver resolution step, and $S \myboxi \varepsilon
\not\red_{\chi}$ indicates that $S$ is in solved form w.r.t. the
action of the constraint solver in the sense of Definition
\ref{defsol}. Moreover, the relation $\red_{\chi}$ manipulates
disjunctions by combining them as follows:
\begin{center}
$\ldots \, \vee S_{i} \myboxi \,\sigma_{i} \vee \, \ldots \,
\red_{\chi} \, \ldots \, \vee  \bigvee_{j} (S_{j} \myboxi \,
\sigma_{j}) \vee \, \ldots$ $~~~~$ if $S_{i} \myboxi
\,\sigma_{i}\red_{\chi} \bigvee_{j} (S_{j} \myboxi \,\sigma_{j})$
\end{center}

Tables~\ref{general rules for FD Hseq}-\ref{Rules for domain and
labeling} show the sets of rules that define the relation
$\red_{\chi}$. These rules extend and complement those presented
in
\cite{fraguas+:lazy-narrowing-calculus-for-declarative-cp-ppdp04}
specifically to work with finite domain constraints defined on the
set of integer patterns. For clarity, we omit the corresponding
failure rules, which can be easily deduced from our tables.

Table~\ref{general rules for FD Hseq} captures the operational
behavior of the constraint solver $Solve^{\cdom}$ to manage
constraints of the form {\em seq}, {\em leq} or {\em domain} when
these return a variable as a result. The result variable is
instantiated to each of its possible values (i.e., $true$ and
$false$) giving rise to different alternatives for each of the
possibilities and propagating the corresponding bind to the
remaining alternatives.

\begin{table}[htbp]
\caption{General Rules for the Constraint Solver} \label{general
rules for FD Hseq}
\begin{center}
\begin{small}
\begin{tabular}{l} \hline\hline
$seq ~t ~s$ $\to!$ $R,$ $S$ $\myboxi$ $\sigma$ $\red_{\chi}$ $(t
== s,$
$S\theta_{1}$ $\myboxi$ $\sigma\theta_{1})$ $\vee$ $(t ~\backslash= s,$
$S\theta_{2}$ $\myboxi$ $\sigma\theta_{2})$ \\
\hline
$leq ~a ~b$ $\to!$ $R,$ $S$ $\myboxi$ $\sigma$ $\red_{\chi}$ $(a \leq b,$
$S\theta_{1}$ $\myboxi$ $\sigma\theta_{1})$ $\vee$ $(a > b,$ $S\theta_{2}$
$\myboxi$ $\sigma\theta_{2})$ \\
\hline
$domain ~u ~[u_{1},\ldots, u_{n}]$ $\to!$ $R,$ $S$ $\myboxi\sigma$ $\red_{\chi}$
$(u \in [u_{1},\ldots, u_{n}],$ $S\theta_{1}$ $\myboxi$ $\sigma\theta_{1})$
$\vee$ \\
\hspace*{5.75cm}$(u \notin [u_{1},\ldots, u_{n}],$ $S\theta_{2}$ $\myboxi$
$\sigma \theta_{2})$ \\\\
(For the 3 rules) only if $R \notin \chi$;  with
$\theta_{1} = \{R \mapsto true\}$ and $\theta_{2} = \{R \mapsto false\}$ \\
\hline\hline
\end{tabular}
\end{small}
\end{center}
\end{table}

Observe that, by applying the rules shown in Table~\ref{general
rules for FD Hseq}, all the constraints based on the primitive
{\em seq}  are proposed as explicit constraints in form of strict
equality or strict disequality. Then, the solver distinguishes
several cases depending on the syntactic structure of the
(integer) patterns used as arguments. Table~\ref{Rules for strict
equality} shows the rules to cover these cases that reproduce the
process of syntactic unification between equalities and
disequalities as it is done in the classical syntactic unification
algorithms (see for example \cite{fernandez:92aaecc}).

\begin{table}[htbp]
\caption{Rules for Strict (Dis)-Equality} \label{Rules for strict
equality}
\begin{center}
\begin{small}
\begin{tabular}{l}
\hline\hline
$u == u,$ $S$ $\myboxi$ $\sigma$ $\red_{\chi}$ $S$ $\myboxi$ $\sigma$,
~if $u\in \mathbb{Z}$\\
$X == t,$ $S$ $\myboxi$ $\sigma$ $\red_{\chi}$ $t == t,$ $S\theta$
$\myboxi$ $\sigma\theta$, ~~if $X$ $\notin$ $\chi$ $\cup$
$var(t),$ $var(t)$ $\cap$ $\chi$ $=$ $\emptyset,$ $\theta$ $=$
$\{X \mapsto t\}$\\
$(h~t_{1}\ldots t_{n}) == (h~s_{1}\ldots s_{n}),$ $S$ $\myboxi$
$\sigma~\red_{\chi}$ $t_{1} == s_{1},$ $\ldots,$ $t_{n} == s_{n},$
$S\myboxi$ $\sigma$\\
\hline
$u ~\backslash= u',$ $S$ $\myboxi$ $\sigma$ $\red_{\chi}$ $S$ $\myboxi$
$\sigma$, ~if $u,u'$ $\in$ $\mathbb{Z},$ and $u$ $\neq^{\mathbb{Z}}$ $u'$  \\
$X ~\backslash= (h~t_{1}\ldots t_{n}),$ $S$ $\myboxi$ $\sigma$
$\red_{\chi}$ $(\bigvee_{i}(S\theta_{i}$ $\myboxi$ $\sigma\theta_{i}))$
$\vee$ $(\bigvee_{k=1}^{n}(U_{k} ~\backslash= t_{k}\theta,$ $S\theta$ $\myboxi$
$\sigma\theta))$   \\
\hspace*{2.cm} if $X$ $\notin$ $\chi,$ $var(h~t_{1}\ldots t_{n})$$\cap$ $\chi$
$\neq$ $\emptyset,$ $\theta_{i}$ $=$ $\{X \mapsto$ $h_{i}$ $\tpp{Y}{m_i}\}$,
with $h_{i}$ $\neq$ $h$, and \\
\hspace*{2.cm} $\theta$ $=$ $\{X \mapsto h$ $\tpp{U}{n}\},$
$\overline{Y}_{m_{i}},$ $\overline{U}_{n}$ are new fresh variables.\\
$(h\ t_{1} \ldots t_{n}) ~\backslash= u,$ $S$ $\myboxi$ $\sigma$
$\red_{\chi}S$ $\myboxi$ $\sigma$ ~if $u$ $\in$ $\mathbb{Z}$\\
$(h~t_{1}\ldots t_{n}) ~\backslash= (h~s_{1}\ldots s_{n}),$ $S$
$\myboxi$ $\sigma$ $\red_{\chi}$ $(t_{1} ~\backslash= s_{1},$ $S$
$\myboxi$ $\sigma)$ $\vee$ $\ldots$
$\vee$ $(t_{n} ~\backslash= s_{n},$ $S$ $\myboxi$ $\sigma)$ \\
$(h~t_{1}\ldots t_{n}) ~\backslash= (h's_{1}\ldots s_{m}),$ $S$
$\myboxi$ $\sigma$ $\red_{\chi}$ $S$ $\myboxi$ $\sigma,$ ~ if $h
\neq h'$ or $m \neq n$ \\
\hline\hline
\end{tabular}
\end{small}
\end{center}
\end{table}

In addition  to the rules for the strict (dis)equality over
integer patterns, the solver has also to consider, by contrast to
the solver given in
\cite{fraguas+:lazy-narrowing-calculus-for-declarative-cp-ppdp04},
new rules for the particular treatment of the primitive
constraints (specific for $\cdom$) defined over the primitive
elements in $\mathbb{Z}$. These rules are shown in Table~\ref{New
rules to complement the solver}.

\begin{table}[htbp]
\caption{Rules for the Specific Primitive Constraints of $\cdom$}
\label{New rules to complement the solver}
\begin{center}
\begin{small}
\begin{tabular}{l} \hline\hline
$u \leq u',$ $S$ $\myboxi$ $\sigma$ $\red_{\chi}$ $S$ $\myboxi$ $\sigma$,
~if $u,u'$ $\in$ $\mathbb{Z}$, and $u$ $\leq^{\mathbb{Z}}$ $u'$ \\
$u > u',$ $S$ $\myboxi$ $\sigma$ $\red_{\chi}$ $S$ $\myboxi$ $\sigma$,
~if $u,u'$ $\in$ $\mathbb{Z}$, and $u$ $>^{\mathbb{Z}}$ $u'$ \\
\hline
$a \otimes b \asymp c,$ $S$ $\myboxi$ $\sigma$ $\red_{\chi}$ $S$ $\myboxi$ $\sigma$,
~if $a, b, c$ $\in$ $\mathbb{Z}$ and $a \otimes^{\mathbb{Z}} b$ $\asymp^{\mathbb{Z}}$ $c$ \\
$a \otimes b = X,$ $S$ $\myboxi$ $\sigma$ $\red_{\chi}$ $S\theta$
$\myboxi$ $\sigma\theta$, ~if $X$ $\notin$ $\chi,$ $a, b$ $\in$
$\mathbb{Z}$ and $\theta$ $=$ $\{X \mapsto a \otimes^{\mathbb{Z}}
b\}$\\
\hline\hline
\end{tabular}
\end{small}
\end{center}
\end{table}

Moreover, the solver also has to cover the {\em domain} and {\em
indomain} classical constraints in finite domain constraint
programming languages, to respectively fix the domain of the
constrained variables and label them according to their
corresponding domain
\cite{dechter:constraint-processing-morgan-kauffmann-2003}.
Table~\ref{Rules for domain and labeling} shows the new rules that
consider these cases.

\begin{table}[htbp]
\caption{Rules for Finite Domain and Variable
Labeling}\label{Rules for domain and labeling}
\begin{center}
\begin{small}
\begin{tabular}{l} \hline\hline
$u$ $\in$ $[u_{1}, \ldots, u_{n}],$ $S$ $\myboxi$ $\sigma$ $\red_{\chi}$ $S$ $\myboxi$ $\sigma$,
~if $u, u_{i}$ $\in$ $\mathbb{Z}$ $\cup$ $\Var$ and
$\exists i$ $\in$ $\{1,\ldots, n\}.$ $u_{i}$ $\equiv$ $u$. \\
$u$ $\notin$ $[u_{1}, \ldots,u_{n}],$ $S$ $\myboxi$ $\sigma$ $\red_{\chi}$ $S$ $\myboxi$ $\sigma$,
~if $u, u_{i}$ $\in$ $\mathbb{Z}$ and
$\forall i$ $\in$ $\{1,\ldots, n\}.$ $u_{i}$ $\neq^{\mathbb{Z}}$ $u$. \\
\hline
$labeling ~[\ldots]~ [X],$ $X$ $\in$ $[u_{1},\ldots, u_{n}],$ $S$ $\myboxi$ $\sigma$ $\red_{\chi}$
$\bigvee_{i=1}^{n} (S\theta_{i}$ $\myboxi$ $\sigma\theta_{i}),$ \\
\hspace*{2.cm}if $X$ $\notin$ $\chi$, and $\forall i$ $\in$ $\{1,\ldots, n\}$,
$u_{i}$ $\in$ $\mathbb{Z}$ and  $\theta_{i}$ $=$ $\{X \mapsto u_{i}\}$ \\
$labeling ~[\ldots] ~ [u],$ $S$ $\myboxi$ $\sigma$ $\red_{\chi}$ $S$ $\myboxi$ $\sigma$,
~if $u$ $\in$ $\mathbb{Z}$ \\
\hline\hline
\end{tabular}
\end{small}
\end{center}
\end{table}

After applying the constraint solver $Solve^{\cdom}$, a primitive
constraint store $S$ $\subseteq$ $PCon(\cdom)$ is expressed in {\em
solved form} as a finite conjunction of primitive constraints of the
form (we use the notations given in Section \ref{domains}) $X == t$,
$X ~\backslash = t$, $u \in D$ and $a \otimes b \asymp c$ where $X
\in \Var$, $t$ is a total pattern, $u, a, b,
c~\in~\mathbb{Z}~\cup~\Var$ and $D$ is a total pattern defining a
list of variables and/or integers.

\begin{example}
We illustrate the operational semantics of our finite domain
constraint solver providing a constraint solver derivation from
the initial constraint store $\{seq~X~(s~K)$ $\to!$ $R,$ $A + B <
Z\}$ and taking into account the set of protected variables
$\{Z\}$. We describe in detail the rules applied by the constraint
solver
and, at each goal transformation step, we underline which subgoal
is selected:\\

$\underline{seq ~X ~(s~K) \to! ~R},$ $A + B < Z$ $\myboxi$
$\varepsilon$ $\red_{\{Z\}}$\\
\hspace*{4.0cm} $(\{X == s~K,$ $A + B < Z\}$ $\myboxi$ $\{R
\mapsto true\})$ $\vee$\\
\hspace*{4.0cm} $(\{X ~\backslash= s~K,$ $A + B < Z\}$ $\myboxi$
$\{R \mapsto false\})$ $\red_{\{Z\}}$
\begin{itemize}
\item $(\{\underline{X == s~K},$ $A + B < Z\}$ $\myboxi$
$\{R \mapsto true\})$ $\red_{\{Z\}}$\\
$(\{\underline{s~K == s~K},$ $A + B < Z\}$ $\myboxi$
$\{R \mapsto true, X \mapsto s~K\})$ $\red_{\{Z\}}$\\
$(\{K == K,$ $A + B < Z\}$ $\myboxi$ $\{R \mapsto true, X \mapsto
s~K\})$ $\not\red_{\{Z\}}$
\item $(\{\underline{X ~\backslash= s~K},$ $A + B < Z\}$ $\myboxi$
$\{R \mapsto false\})$ $\red_{\{Z\}}$\\
\hspace*{2.0cm}$(\{A + B < Z\}$ $\myboxi$
$\{R \mapsto false,$ $X \mapsto 0\})$ $\vee$\\
\hspace*{2.0cm}$(\{A + B < Z,$ $M ~\backslash= K\}$ $\myboxi$ $\{R
\mapsto false,$ $X \mapsto s~M\})$ $\not\red_{\{Z\}}$
\end{itemize}

Therefore, the constraint solver returns the following solved forms:
\\

$Solve^{\cdom}(\{$ $seq~X~(s~K)$ $\to!$ $R,$ $A + B < Z$ $\},$ $\{Z\}$ $)$ $=$\\
\hspace*{4.0cm} $(\{A + B < Z,$ $K == K\}$ $\myboxi$ $\{R \mapsto
true,$ $X \mapsto s~K\})$ $\vee$\\
\hspace*{4.0cm} $(\{A + B < Z\}$ $\myboxi$ $\{R \mapsto false,$ $X \mapsto 0\})$ $\vee$\\
\hspace*{4.0cm} $(\{A + B < Z,$ $M ~\backslash= K\}$ $\myboxi$
$\{R \mapsto false,$ $X \mapsto s~M\})$
\end{example}

As shown in Tables 2-5, our new constraint solver for the finite
domain $\cdom$ with strict equality and disequa\-lity has been
designed to hold all the initial assumptions required in the general
framework $\Cflp$ for constraint solvers (see Definition
\ref{defsol}). It can be formally proved by means of the following
result.

\newtheorem{theorem}{Theorem}\label{teorema}
\begin{theorem}
Let $S$ $\subseteq$ $PCon(\cdom)$ be a primitive constraint store,
$\sigma$ $\in$ $\sub{\mathbb{Z}}$ an idempotent total substitution
and $\chi$ $\subseteq$ $\Var$ a set of protected variables. ~If
$S~\myboxi~\sigma$ satisfies the requirements of Definition
\ref{defsol} ~and~ $S~\myboxi~\sigma~\red_{\chi}~\bigvee_{i =
1}^{k}(S_i~\myboxi~\sigma_i)$, ~then~
$Sol_{\cdom}(S~\myboxi~\sigma)~=~$ $\bigcup_{i = 1}^{k}$
$Sol_{\cdom}(S_i~\myboxi~\sigma_i)$, where $dom(\sigma_i)$ $\cap$
$var(S_i)~=~\emptyset$ and
$\chi~\cap~(dom(\sigma_i)~\cup~ran(\sigma_i))$ $=~\emptyset$ for all
$1 \leq i \leq k$. Moreover, if $S~\myboxi~\sigma$ $\red_{\chi}$
fails then $Sol_{\cdom}(S~\myboxi~\sigma)~=~\emptyset$.
\end{theorem}

The proof of this theorem (see \ref{appA}) can be done
distinguishing several cases from the declarative semantics of
each primitive function symbol given in Table 1 and the
requirements of each constraint solver rule in Tables 2-5.
According to this result, the relation $\red_{\chi}$ preserves the
requirements of a constraint solver and the constraint solver
steps fail only in case of an unsatisfiable constraint store.
Therefore, if we repeatedly apply this result from an initial
constraint store and a set of protected variables in order to
compute a constraint store in solved form, we directly obtain the
correctness of our finite domain constraint solver w.r.t.
$\cflp{\cdom}$'s semantics.

\section{The $CLNC(\cdom)$ Calculus} \label{sect: object}
This section describes a lazy narrowing calculus with constraints
defined on the finite domain $\cdom$ (the Constraint Lazy
Narrowing Calculus $CLNC(\cdom)$ for short) for the solving of
goals from programs. Since we have proved in the previous section
that our finite domain constraint solver holds the properties
required in the general framework, this calculus can be obtained
as a simplified instantiation of the general scheme for $\Cflp$
described in
\cite{fraguas+:lazy-narrowing-calculus-for-declarative-cp-ppdp04},
and used in this work as the formal foundation of the operational
semantics in $\toyfd$.

In order to understand the main ideas of our operational
semantics, we first give a precise definition for the class of
well-typed programs, admissible goals and correct answers we are
going to work with.

\subsection{Programs, Goals and Answers} \label{programs}

Our well-typed {\em $\cflp{\cdom}$-programs} are sets of
constrained rewriting rules that define the behavior of possibly
higher-order and/or non-deterministic lazy functions over $\cdom$,
called {\em program rules}. More precisely, a program rule $R$ for
a defined function symbol $f \in \df{n}$ with an associated
principal type $\tau_{1} \to \ldots \to \tau_{n} \to \tau$ has the
form $f\, t_{1} \ldots t_{n}\, = r\,\Leftarrow C$ and is required
to satisfy the conditions listed below:
\begin{enumerate}
\item $t_1 \dots t_n$ is a linear sequence of transparent patterns
and $r \in \expr{\mathbb{Z}}$ is a total expression. \item $C$ is a
finite set of total constraints (in the form of Definition
\ref{semantica}), intended to be interpreted as conjunction, and
possibly including occurrences of defined function symbols. \item
There exists some type environment $T$ with domain $var(R)$ which
    well-types the defining rule in the following sense:
\begin{enumerate}
    \item For $1 \leq i \leq n$:  ${(\Sigma, T) \vdash_{WT}\ }t_i~::~\tau_i$.
    \item  ${(\Sigma, T) \vdash_{WT}\ }r~::~\tau$.
    \item  For each $(e == e') \in C$: $\exists \mu \in \Type$ s.t.
    ${(\Sigma, T) \vdash_{WT}\ } e~::~\mu~::~e'$.
    \item  For each $(e~ \backslash = e') \in C$: $\exists \mu \in \Type$ s.t.
    ${(\Sigma, T) \vdash_{WT}\ } e~::~\mu~::~e'$.
    \item  For each $(u\in D)\in C$: ${(\Sigma, T)\vdash_{WT}\ }
    u~::~int, D~::~[int]$.
    \item  For each ($a \otimes b\asymp c) \in C$:
    ${(\Sigma, T)\vdash_{WT}\ } a, b, c~::~int$
\end{enumerate}
where ${(\Sigma, T) \vdash_{WT}\ } e :: \tau :: e'$ denotes
${(\Sigma, T) \vdash_{WT}\ } e :: \tau,\ {(\Sigma, T) \vdash_{WT}\
}e' :: \tau$.
\end{enumerate}

\hspace*{2mm} The left-linearity condition required in item $1$ is
quite common in functional and functional logic programming. As in
constraint logic programming, the conditional part of a program
rule needs no explicit occurrences of existential quantifiers.
Another distinguished feature of our language is that no
confluence properties are required for the programs, and therefore
functions can be {\em non-deterministic}, i.e. return several
values for given (even ground) arguments.

\begin{example} \label{lazy}
The following example illustrates the previous definition of typed
$\cflp{\cdom}$-programs by showing some constrained program rules
which will be used for lazy evaluation of infinite lists in the
next subsections.\\

{\em
take :: int $\to$ $[\alpha]$ $\to$ $[\alpha]$ \\
\hspace*{-1.5cm}
\begin{tabular}{lll}
{\bf (T1)}  take $0$  Xs            & =  $[\,]$ & \\
{\bf (T2)}  take  N  $[\,]$         & =  $[\,]$ & $\Leftarrow$ $N > 0$ \\
{\bf (T3)} take  N  $[$X $\mid$ Xs$]$  & = $[$X$\mid$take~(N - 1)~
Xs$]$ & $\Leftarrow$
N $> 0$ \\\\
\end{tabular}
}

{\em
check\_list :: $[$int$]$ $\to$ int \\
\hspace*{-2.cm}
\begin{tabular}{lll}
{\bf (CL1)} check\_list $[\,]$      & =  $0$ &  \\
{\bf (CL2)} check\_list $[$X$\mid$Xs$]$ & =  $1$ & $\Leftarrow$ domain~$[$X$]$~1~2 \\
{\bf (CL3)} check\_list $[$X$\mid$Xs$]$ & =  $2$ & $\Leftarrow$ domain~$[$X$]$~3~4 \\
{\bf (CL4)} check\_list $[$X$\mid$Xs$]$ & =  $4$ & $\Leftarrow$
domain~$[$X$]$~5~7 \\\\
\end{tabular}
}

{\em generateFD :: int $\to$ $[$int$]$ \\
\begin{tabular}{lll}
{\bf (G1)} generateFD $0$ & = $[~]$ &\\
{\bf (G2)} generateFD  N  & = $[$X $\mid$ generateFD ~N$]$ &
$\Leftarrow$ N $> 0$, domain~$[$X$]$~0~N-1\\\\
\end{tabular}
}

{\em from :: int $\to$ $[$int$]$\\
\hspace*{.5cm}{\bf (F)} from~ N   = $[$N $\mid$ from~(N+1)$]$ }
\end{example}

According to
\cite{fraguas+:lazy-narrowing-calculus-for-declarative-cp-ppdp04},
we define {\em goals} for this kind of programs in the general form
$G \equiv \ex \utup.$ $P$ $\myboxi$ $C$ $\myboxi$ $S$ $\myboxi$
$\sigma$, where the symbol $\myboxi$ must be interpreted as
conjunction, $\utup$ is the finite set of so-called {\em existential
variables} of the goal $G$, $P$ is a multiset of so-called {\em
productions} of the form $e_1 \to t_1,\ldots,e_n \to t_n$, where
$e_i \in \expr{\mathbb{Z}}$ and $t_i \in \pat{\mathbb{Z}}$ are
totals for all $1 \leq i \leq n$ (the set of {\em produced
variables} of $G$ is defined as the set of variables occurring in
$t_1 \ldots t_n$), $C$ is a finite conjunction of constraints
(possibly including occurrences of defined function symbols), $S$ is
a finite conjunction of primitive constraints in the form of
Definition \ref{semantica}, called {\em constraint store}, and
$\sigma$ is an idempotent substitution called {\em answer
substitution} such that $dom(\sigma)$ $\cap$ $var(P$ $\myboxi$ $C$
$\myboxi$ $S)$ $=$ $\emptyset$.

Additionally, we say that a goal $G$ is an {\em admissible goal} iff
it is well-typed: satisfies the same admissibility criteria given
above for programs for each constraint in $C$ and $S$,  and the same
conditions of compatible types for each production in $P$ and each
binding in $\sigma$ given in
\cite{gonzalez+:polymorphic-types-FLP-flops99}. Moreover, it must
hold the so-called {\em goal invariants} given in
\cite{fraguas+:lazy-narrowing-calculus-for-declarative-cp-ppdp04}:
each produced variable is produced only once, all the produced
variables must be existential, the transitive closure of the
relation between produced variables must be irreflexive, and no
produced variable enters the answer substitution. An admissible goal
is called a {\em solved goal} iff $P$ and $C$ are empty and $S$ is
in solved form w.r.t. the action of the constraint solver in the
sense of Definition \ref{defsol}.

Similarly to
\cite{gonzalez+:declarative-rewriting-logic-jlp99,gonzalez+:polymorphic-types-FLP-flops99,vado+:ppdp03},
the $\clnc{\cdom}$ calculus uses a notion of {\em demanded
variable} to deal with lazy evaluation.

\begin{definition}

Let $G$ be an admissible goal. We say that $X$ $\in$ $var(G)$ is a
{\em demanded variable} iff
\begin{enumerate}
\item $X$ is {\em demanded by the constraint store} $S$ of $G$,
i.e. $\mu(X) \neq \bot$ holds for every $\mu \in Sol_{\cdom}(S)$
(for practical use, the calculation of this kind of demanded
variables in $\cdom$ can be easily done extending the rules given in
the appendix of
\cite{fraguas+:lazy-narrowing-calculus-for-declarative-cp-ppdp04} in
the line of our rules shown in Tables 2-5).
\item $X$ is {\em demanded by a production} $(X \tpp{a}{k}
\rightarrow t) \in P$ such that, $t \notin \Var$ or $k
> 0$ and $t$ is a demanded variable in $G$.
\end{enumerate}

\end{definition}

\begin{example}
We suppose an admissible goal with only the primitive constraint
$seq$ $X$ $(s~K)$ $\to!$ $R$ in the associated constraint store
$S$. We note that $K$ is not a demanded variable by $S$, because
$\mu$ $=$ $\{X$ $\mapsto$ $0,$ $K$ $\mapsto$ $\bot,$ $R$ $\mapsto$
$false\}$ $\in$ $Sol_{\cdom}(S)$ (clearly, $seq^{\cdom}$ $\mu(X)$
$(s~K)\mu$ $\to$ $\mu(R)$ $=$ $false$ where $\mu(X)$ $=$ $0$ and
$(s~K)\mu$ $=$ $s~(\mu(K))$ $=$ $s~(\bot)$ have no common upper
bound w.r.t. the information ordering $\sqsubseteq$, according to
Table $1$) but $\mu(K)$ $=$ $\bot$. However, $X$ and $R$ are both
demanded variables by $S$ (according to the {\em radicality}
property, any $\mu$ $\in$ $Sol_{\cdom}(S)$ must satisfy $\mu(R)$
total and then $\mu(R)$ $\neq$ $\bot$ and consequently $\mu(X)$
$\neq$ $\bot$). In this situation, if we have also a production
$F~1$ $\to$ $X$ in the produced part of the goal involving a
higher-order variable $F$, automatically $F$ is also a demanded
variable (by a production but not by the constraint store $S$).
Moreover, we note that it is also possible to have a variable $F$
demanded by both the constraint store (for example, if we add the
primitive constraint $F$ $==$ $\oplus~2$ to $S$) and a production
(for example, $F~1$ $\to$ $3$ instead of $F~1$ $\to$ $X$). In this
case, $F$ is demanded twice, supplying more relevant and precise
information for goal solving in the produced part and the
constraint store of the goal.
\end{example}

Finally, we describe the notion of correct answer that we want to
compute from goals and programs in our $\cflp{\cdom}$-framework.
Since the calculus $CLNC(\cdom)$ is semantically based on the {\em
Constraint ReWriting Calculus} $CRWL(\cdom)$, that represents a
concrete instance over the constraint domain $\cdom$ of the
constraint rewriting logic described in
\cite{fraguas+:constraint-functional-logic-progr-rev-wrla04}, this
logic can be also used as a logical characterization of our
program semantics. On the basis of this logic, we define our
concept of correct {\em answer} with respect to an admissible goal
$G$ and a given $\cflpfd$-program as a pair of the form $\Pi$
$\myboxi$ $\theta$, where $\Pi$ $\subseteq$ $PCon(\cdom)$ and
$\theta$ $\in$ $Sub_{\perp}(\mathbb{Z})$ is an idempotent
substitution such that $dom(\theta)$ $\cap$ $var(\Pi)$ $=$
$\emptyset$, fulfilling the same semantic conditions given in
\cite{fraguas+:lazy-narrowing-calculus-for-declarative-cp-ppdp04}
w.r.t. $\crwl{\mathcal{D}}$'s semantics.

The following example shows a correct answer for the admissible goal
with only a strict equality {\bf $\myboxi$ $take$ $3$ $(generateFD$
$10)$ $==$ $List$ $\myboxi$ $\myboxi$} and the $\cflpfd$-program
given in Example \ref{lazy}:
\begin{center}
{\bf $\{X_{1},X_{2},X_{3}$ $\in$ $[0,1,2,3,4,5,6,7,8,9]\}$ $\Box$
$\{List$ $\mapsto$ $[X_1,X_2,X_3]\}$}
\end{center}

Analogously, it is also possible to prove that {\bf $M \in [1..2]$}
and {\bf $M \in [3..4]$} (both of them with an empty substitution)
are correct answers for the admissible goal with only a user
defined finite domain constraint {\bf $\myboxi$ $check\_list$
$(from$ $M)$ $<$ $3$ $\myboxi$ $\myboxi$}. We will see in the next
subsection how to compute all of these answers by means of the
constrained lazy narrowing over $\cdom$.
\subsection{Constrained Lazy Narrowing over $\cdom$} \label{productions}

The calculus $\clnc{\cdom}$ can be obtained as a particular
instantiation from the general $CLNC(\mathcal{D})$ calculus because
we have proved that our finite domain constraint solver satisfies
the requirements given in the general framework. Therefore, the
calculus $CLNC(\cdom)$ can be described as a set of transformation
rules for admissible goals of the form $G \red G'$, specifying one
of the possible ways of performing one step of goal solving. In this
sense, {\em derivations} are sequences of $\red$-steps where
successful derivations will eventually end with a solved goal and
failing derivations end with an inconsistent goal $\blacksquare$. We
have two classes of goal transformation rules: rules for constrained
lazy narrowing by means of productions, and rules for constraint
solving and failure detection.

The goal transformation rules concerning productions are the same
rules given in
\cite{fraguas+:lazy-narrowing-calculus-for-declarative-cp-ppdp04}
for general productions and are designed with the aim of modeling
the behavior of constrained lazy narrowing with {\em sharing}, but
now involving only the primitive functions over finite domains given
in Definition \ref{semantica}, possibly higher-order defined
functions and functional variables.

The goal transformation rules concerning constraints can also be
used to combine (primitive or used defined) finite domain
constraints with the action of our constraint finite domain
solver. As the main novelty, we note that only primitive
constraints are sent to the ${\cal FD}$ constraint solver. This is
because non-primitive constraints are first translated to
primitive ones by replacing the non-primitives arguments by new
fresh variables before executing constraint solving and by
registering new productions between the non-primitive arguments
and the new variables for lazy evaluation. Moreover, the
constraint solver must protect all the produced variables of the
goal in order to respect the constrained lazy evaluation and the
admissibility conditions of goals. Additionally, the usual failure
rules can also be used for failure detection in constraint solving
and failure detection in the syntactic unification of the produced
part of the goal.

Finally,  we note that since Theorem 1 proves the correctness of
our finite domain constraint solver w.r.t. the general framework,
the main properties of the lazy narrowing calculus $\clnc{\cdom}$,
{\em soundness} and {\em completeness} w.r.t. the declarative
semantics of $\crwl{\cdom}$, follows directly from the general
results of
\cite{fraguas+:lazy-narrowing-calculus-for-declarative-cp-ppdp04}.
Obviously, these properties qualify $\clnc{\cdom}$ as a convenient
computation mechanism for constraint functional logic programming
over finite domains and provide a formal foundation for our
$\cflp{\cdom}$ implementation $\toyfd$. From the viewpoint of
efficiency, a computation strategy for $\clnc{\cdom}$ using {\em
definitional trees} \cite{antoy:definitional-trees-alp-92} has
been proposed recently in \cite{vado+:frocos05} and
\cite{soniadelvado+:wcflp05} for ensuring only needed narrowing
steps and extend the efficient properties shown in
\cite{DBLP:conf/plilp/LoogenFR93,hanus:+needed,vado+:ppdp03}
guiding and avoiding {\em don't know} choices of constrained
program rules over $\cdom$.

\subsection{Example of Goal Resolution by Using $CLNC(\cdom)$}
\label{example of goal resolution}

This section is closed with a simple example which illustrates the
process of goal solving via the narrowing calculus $\clnc{\cdom}$
and our finite domain constraint solver $Solve^{\cdom}$. We compute
all the answers from the goal $\myboxi$ $check\_list$ $(from$ $M)$
$<$ $3$ $\myboxi$ $\myboxi$ using the $\cflp{\cdom}$-programs given
in Example \ref{lazy}. Its resolution corresponds to the following
sequence of goal transformation rules in
\cite{fraguas+:lazy-narrowing-calculus-for-declarative-cp-ppdp04}
where, at each goal transformation step, we underline which subgoal
is selected. $\red_{{\bf R}}$ indicates that the rule ${\bf R}$ in
that work is applied.
\\\\
$\myboxi$ $\underline{check\_list~(from~M)~<~3}$ $\myboxi$
$\myboxi$ $\varepsilon$ $\red_{{\bf AC}}$\\
$\exists X.$ $\underline{check\_list~(from~M) \to X}$ $\myboxi$
$\myboxi$ $X
< 3$ $\myboxi$ $\varepsilon$ $\red_{{\bf DF}}$\\\\
At this point, we note that $X$ is a variable demanded by the
constraint store and we have several alternatives due to {\em don't
know} choice of
the program rule $check\_list$:\\\\
$\exists X.$ $\underline{check\_list~(from~M) \to X}$ $\myboxi$
$\myboxi$ $X
< 3$ $\myboxi$ $\varepsilon$ $\red_{{\bf DF(CL1)}}$\\
$\exists X.$ $from~M \to [\,],$ $\underline{0 \to X}$ $\myboxi$
$\myboxi$ $X
< 3$ $\myboxi$ $\varepsilon$ $\red_{{\bf SP\{X \mapsto 0\}}}$\\
$from~M \to [\,]$ $\myboxi$ $\myboxi$ $\underline{0 < 3}$
$\myboxi$ $\varepsilon$ $\red_{{\bf CS\{\emptyset\}}}$
$~~~~(Solve^{\cdom}(\{0 < 3\},\emptyset)$ $=$ $\emptyset$
$\myboxi$
$\varepsilon$)\\
$\underline{from~M \to [\,]}$ $\myboxi$ $\myboxi$ $\myboxi$ $\varepsilon$ $\red_{{\bf DF(F)}}$\\
$\underline{[M~|~from~(M+1)] \to [\,]}$ $\myboxi$ $\myboxi$
$\myboxi$ $\varepsilon$ $\red_{{\bf CF}}$
$\blacksquare$\\\\
The application of the first program rule for $check\_list$ leads
to a failure derivation without answer. We apply now the second
program rule of $check\_list$:\\\\
$\exists X.$ $\underline{check\_list~(from~M) \to X}$ $\myboxi$
$\myboxi$ $X
< 3$ $\myboxi$ $\varepsilon$ $\red_{{\bf DF(CL2)}}$\\
$\exists X',Xs',X.$ $from~M \to [X'|Xs'],$ $\underline{1 \to X}$
$\myboxi$\\
\hspace*{6.0cm} $domain~[X']~1~2$ $\myboxi$ $X
< 3$ $\myboxi$ $\varepsilon$ $\red_{{\bf SP\{X \mapsto 1\}}}$\\
$\exists X',Xs'.$ $from~M \to [X'|Xs']$ $\myboxi$
$\underline{domain~[X']~1~2}$ $\myboxi$ $1 < 3$ $\myboxi$
$\varepsilon$ $\red_{{\bf AC}}$\\
$\exists X',Xs'.$ $\underline{from~M \to [X'|Xs']}$ $\myboxi$
$\myboxi$ $1 < 3,$
$domain~[X']~1~2$ $\myboxi$ $\varepsilon$ $\red_{\bf DF(F)}$\\
$\exists X',Xs'.$ $\underline{[M~|~from~(M+1)] \to [X'|Xs']}$
$\myboxi$ $\myboxi$ $1 < 3,$
$domain~[X']~1~2$ $\myboxi$ $\varepsilon$ $\red_{\bf DC}$\\
$\exists X',Xs'.$ $\underline{M \to X'},$ $from~(M+1) \to Xs'$
$\myboxi$ $\myboxi$\\
\hspace*{6.0cm} $1 < 3,$
$domain~[X']~1~2$ $\myboxi$ $\varepsilon$ $\red_{\bf SP\{X' \mapsto M\}}$\\
$\exists Xs'.$ $\underline{from~(M+1) \to Xs'}$ $\myboxi$
$\myboxi$ $1 < 3,$
$domain~[M]~1~2$ $\myboxi$ $\varepsilon$ $\red_{\bf EL}$\\
$\myboxi$ $\myboxi$ $\underline{1 < 3, domain~[M]~1~2}$ $\myboxi$
$\varepsilon$ $\red_{\bf CS(\emptyset)}$ $\myboxi$ $\myboxi$ $M
\in [1..2]$ $\myboxi$ $\varepsilon$, because\\
$Solve^{\cdom}(\{1 < 3,domain~[M]~1~2\},\emptyset)$ $=$ $\{M \in
[1..2]\}$ $\myboxi$
$\varepsilon$\\\\
Therefore, we obtain the first computed answer $\Pi_1$ $\myboxi$
$\theta_1$ $\equiv$ $\{M \in [1..2]\}$ $\myboxi$ $\varepsilon$.
Analogously, we can apply the third program rule of $check\_list$:
\vspace*{.15cm}

\hspace*{-.3cm}$\exists X.$ $\underline{check\_list~(from~M) \to X}$ $\myboxi$
$\myboxi$ $X
< 3$ $\myboxi$ $\varepsilon$ $\red^{*}_{{\bf DF(CL3)}}$\\
$\myboxi$ $\myboxi$ $M \in [3..4]$ $\myboxi$ $\varepsilon$
\vspace*{.15cm}

\noindent and we obtain the second computed answer $\Pi_2$
$\myboxi$ $\theta_2$ $\equiv$ $\{M \in [3..4]\}$ $\myboxi$
$\varepsilon$. No more answers can be computed, because if we
apply the fourth
program rule of $check\_list$ we have again a failing derivation:\\\\
$\exists X.$ $\underline{check\_list~(from~M) \to X}$ $\myboxi$
$\myboxi$ $X
< 3$ $\myboxi$ $\varepsilon$ $\red_{{\bf DF(CL4)}}$\\
$\exists X',Xs',X.$ $from~M \to [X'|Xs'],$ $\underline{4 \to X}$
 $\myboxi$ $domain~[X']~5~7$ $\myboxi$ $X
< 3$ $\myboxi$ $\varepsilon$ $\red_{{\bf SP\{X \mapsto 4\}}}$\\
$\exists X',Xs'.$ $from~M \to [X'|Xs']$ $\myboxi$
$\underline{domain~[X']~5~7}$ $\myboxi$ $4
< 3$ $\myboxi$ $\varepsilon$ $\red_{{\bf AC}}$\\
$\exists X',Xs'.$ $from~M \to [X'|Xs']$ $\myboxi$ $\myboxi$
$\underline{4 < 3, domain~[X']~5~7}$ $\myboxi$ $\varepsilon$
$\red_{{\bf SF\{X',Xs'\}}}$ $\blacksquare$\\
because $Solve^{\cdom}(\{4 < 3,domain~[X']~5~7\},\{X',Xs'\})$ $=$
$\emptyset$

\vspace{3mm}

A detailed explanation of the computation of these answers using
definitional trees in $\clnc{\cdom}$ to efficiently guide and
avoid {\em don't know} choices of constrained program rules can be
found in \cite{soniadelvado+:wcflp05}. Moreover, we will see in
the next section that these are exactly the same answers computed
by our $\cflp{\cdom}$ implementation $\toyfd$.

%

\section{$\Toyfd$}\label{sect:TOYFD}

So far, we have introduced the theoretical framework. Now, in this
section we introduce $\toyfd$, a $\Cflpfd$ implementation that
extends the $\Toy$ system to deal with ${\cal FD}$ constraints,
highlight its advantages, and show its performance.

\subsection{Introducing $\toyfd$}\label{sect:implementation issues}

In this section, we describe $\toyfd$ from a programming point of
view, briefly describing its concrete syntax and some predefined
${\cal FD}$ constraints.

\subsubsection{An Overview of $\toyfd$} \label{an overview of toyfd}

\begin{sloppypar}
$\Toyfd$ programs consist of {\em datatypes}, {\em type alias},
{\em infix operator} definitions, and rules for defining {\em
functions}. The syntax is mostly borrowed from Haskell with the
remarkable exception that variables and type variables begin with
upper-case letters, whereas constructor symbols and type symbols
begin with lower-case. In particular, functions are {\em curried}
and the usual conventions about associativity of application hold.
As usual in functional programming, types are inferred, checked
and, optionally, can be declared in the program. To illustrate the
datatype definitions, we present the following examples using the
concrete syntax of $\Toy$:
\begin{itemize}
    \item {\tt data nat = zero $|$ suc nat}, to define the naturals, and
    \item  the Boolean predefined type  as {\tt data bool = false $|$ true};
\end{itemize}
\end{sloppypar}

A $\Toyfd$ program $P$ is  a set of {\em defining rules} for the
function symbols in its signature. Defining rules for a function
$f$ have the syntactic basic form  $f ~t_1 ~\ldots ~ t_n = r <==
C$ and, informally, its intended meaning is that a call to $f$ can
be reduced to $r$ whenever the actual parameters match the
patterns $t_i$, and the conditions in $C$ are satisfied. $\Toyfd$
also allows predicates (defined similarly as in logic programming)
where predicates are viewed as a particular kind of functions,
with type $p~::~\overline{\tau}_n \to$ {\tt bool}. As a syntactic
facility, we can use {\em clauses} as a shorthand for defining
rules whose right-hand side is $true$. This allows to write
Prolog-like predicate definitions, so that a clause $p ~t_1
~\ldots ~ t_n ~:-~C$ abbreviates a defining rule of the form $p
~t_1 ~\ldots ~ t_n = true <== C$. With this sugaring in mind and
some obvious changes (like currying elimination) it should be
clear that (pure) $\Clpfd$-programs can be straightforwardly
translated to $\Cflpfd$-programs.

\subsubsection{Simple Programming Examples}

Table~\ref{programming-example} shows introductory programming
examples in $\Toy$ that do not make use of the extension over
${\cal FD}$, together with some goals and their outcomes
\cite{fraguas+:multiparadigm-system-rta99}. Note that infix
constraint operators are allowed in $\Toyfd$, such as ~ {\tt //} ~
to build the expression ~ {\tt X // Y}, ~ which is understood as ~
{\tt // X~Y}. The goal {\em (a)} in the table sorts a list, in a
pure functional computation. The answer for the goal {\em (b)}
involves a syntactic disequality. In goal {\em (c)}, {\tt F} is a
higher-order logic variable, and the obtained values for this
variable are higher-order patterns ({\tt permut}, {\tt sort},...).

\begin{table}[htbp]
\caption{$\Toy$ Programming Basic
Examples}\label{programming-example}
\begin{center}
\begin{tabular}{l}
\hline\hline
   \begin{tabular}{l}
    {\sl \% Non-deterministic choice of one of two values}\\
    \qquad {\tt infixr 40 //}\\
    \qquad {\tt X // Y = X}\\
    \qquad {\tt X // Y = Y}\\
    {\sl \% Non-deterministic insertion of an element into a list}\\
    \qquad {\tt insert X [] = [X]} \\
    \qquad {\tt insert X [Y|Ys] = [X,Y|Ys] // [Y|insert X Ys]}\\
    {\sl \% Non-deterministic generation of list permutations}\\
    \qquad  {\tt permut [] = []} \\
    \qquad {\tt permut [X|Xs] = insert X (permut Xs)}\\
    {\sl \% Testing whether a list of numbers is sorted}\\
    \qquad {\tt sorted [] = true} \\
    \qquad {\tt sorted [X] = true} \\
    \qquad {\tt sorted [X,Y|Ys] = sorted [Y|Ys] <== X <= Y}\\
{\sl\% Lazy `generate-and-test' permutation sort. 'check' calls 'sorted' which demands its} \\
{\sl\% argument, which is lazily, non-deterministically generated by
'permut'. As soon as} \\
{\sl\% the test fails, 'permut' stops the generation and tries another
alternative}\\
    \qquad {\tt sort Xs = check (permut Xs)} \\
    \qquad {\tt check Xs = Xs <== sorted Xs == true}
      \end{tabular}\\
      \hline
      \begin{tabular}{ll}
        \hspace{1cm} {\em Goal} \hspace{5cm} &{\em Answers}\\
      \end{tabular}\\
      \hline
      \begin{tabular}{ll|ll}
       {\em (a)} {\tt sort [4,2,5,1,3] == L} & & &
       {\tt L == [1,2,3,4,5]; no more solutions}\\ {\em (b)} {\tt sort\ [3,2,1] /= L}
        & & &{\tt L /= [1,2,3] ; no more  solutions}\\
      {\em (c)} {\tt F [2,1,3] == [1,2,3]}  & & &{\tt F == permut; F
       == sort; $\ldots$} \\
      \end{tabular}\\
      \hline\hline
    \end{tabular}
\end{center}
\end{table}

\subsubsection{${\cal FD}$ Constraints in $\toyfd$}\label{FD constraints in TOY(FD)}

Table~\ref{table:FD constraint set for toy(FD)} shows a small
subset of the ${\cal FD}$ constraints supported by $\Toyfd$, which
are typical instances found in $CP$ systems, and covers adequately
the primitive constraints summarized in Table~\ref{set of
primitive constraints}. In Table~\ref{table:FD constraint set for
toy(FD)}, {\tt int} is a predefined type for integers, and {\tt
[$\tau$]} is the type `list of $\tau$'. The datatype {\tt
labelType} is a predefined type which is used to define many
search strategies for finite domain variable labeling
\cite{fernandez+:toy(fd)-manual02}.

\begin{table}[htbp]
\caption{A Subset of Predefined ${\cal FD}$ Constraints in
$\toyfd$} \label{table:FD constraint set for toy(FD)} {\tt
\begin{tabular}{ll}\hline\hline
RELATIONAL CONSTRAINT OPERATORS \\
    \qquad $(\#=)$, $(\#\backslash\!=)$ :: int $\rightarrow$ int $\rightarrow$ bool & {\bf (Strict Equality)}\\
    \qquad $(\#\!\!<)$, $(\#\!\!< =)$, $(\#\!\! >)$, $(\#\!\!> =)$ :: int $\rightarrow$ int $\rightarrow$ bool & {\bf (Less or Equal)}\\
ARITHMETICAL CONSTRAINT OPERATORS \\
    \qquad $(\#+)$, $(\#-)$, $(\#*)$, $(\#/)$ :: int $\rightarrow$ int $\rightarrow$ int  & {\bf (Operators)}\\
MEMBERSHIP CONSTRAINTS \\
    \qquad domain :: [int] $\rightarrow$ int $\rightarrow$ int $\rightarrow$ bool  & {\bf (Finite Domains)}\\
ENUMERATION CONSTRAINTS \\
    \qquad labeling :: [labelType] $\rightarrow$ [int] $\rightarrow$ bool  & {\bf (Variable Labeling)}\\
COMBINATORIAL CONSTRAINTS \\
    \qquad  all\_different :: [int] $\rightarrow$ bool  & {\bf (Global Constraints)}\\
\hline\hline
\end{tabular}
}
\end{table}

Relational constraint operators are applied to integers and return a
Boolean value. Arithmetical constraint operators are applied to and
return integer values (the set of primitive elements). They can be
combined with relational constraint operators to build (non)linear
(dis)equations as constraints. Moreover, reified
constraints\footnote{{\em Reified constraints} reflect the
entailment of a constraint in a Boolean variable. In general,
constraints in {\em reified} form allow their fulfillment to be
reflected back in an $\cdom$ variable. For example, $X = (Y + Z >
V)$ constrains $X$ to $true$ as soon as the disequation is known to
be true and to $false$ as soon as the disequation is known to be
false. On the other hand, constraining $X$ to $true$ imposes the
disequation, and constraining $X$ to $false$ imposes its negation.
Usually, in $\Clpfd$ languages, the Boolean values $false$ and
$true$ directly correspond to the numerical values 0 and 1,
respectively.} can be implemented by equating a Boolean variable to
a Boolean constraint, for all of the constraints built from the
operators in this table and the contraint {\tt domain} (see
Example~\ref{ex: combination relational-equality constraints}). Due
to the functional component, we can apply this technique to equate
Boolean expressions to Boolean constraints, as well. Both relational
and arithmetical constraint operators are syntactically
distinguished (by prefixing them with $\#$) from standard relational
operators in order to denote its different operational behavior.
Whereas a standard arithmetical operator demands its arguments, an
arithmetical constraint does not. The membership constraint {\tt
domain} restricts a list of variables (its first argument) to have
values in an integer interval (defined by its two next integer
arguments) whenever its return value is {\tt true}, whereas it
restricts these variables to have values different from the interval
when its return value is {\tt false}. The enumeration constraint
{\tt labeling} assigns values to the variables in its input integer
list according to the options specified with the argument of type
list of {\tt labelType}. In this list, search strategies, such as
{\em first-fail} (see Section~\ref{labelings}), as well as
optimization options for finding minimum and maximum values for cost
functions can be specified. The combinatorial constraint {\tt
all\_different} ensures different values for the elements in its
list argument and is an example of the set of global constraints
(for which an efficient propagation algorithm has been developed)
supported by $\Toyfd$.

We do neither mention nor explain all the predefined constraints
in detail and encourage the interested reader to visit the link
proposed in \cite{fernandez+:toy(fd)-manual02} for a more detailed
explanation. We emphasize that all the pieces of code in this
paper are executable in $\toy({\cal FD})$ and the answers for
example goals correspond to actual executions of the programs.

%
%

\subsubsection{Simple Examples with ${\cal FD}$ Constraints}
\label{n-queens problem example}

\begin{example}\label{ex: combination relational-equality constraints}
Below, we show the resolution at the $\Toyfd$ command line level
of a simple goal that does not involve labeling.

\begin{verbatim}
TOY(FD)> domain [X, Y] 10 20, X #<= Y == L
         yes     L == true, X in 10..20, Y in 10..20;
         yes     L == false, X in 11..20, Y in 10..19;
         no
\end{verbatim}

\begin{sloppypar}
Also note that this $\Cflpfd$ implementation only inform about a
limited outcome, which consists of: (1) substitutions of the form
{\tt Variable == Pattern}, (2) disequality constraints {\tt
Variable /= Pattern}, (3) disjunctions $D$ of constraints {\tt
Variable in IntegerRange} (these constraints denote the possible
values a variable {\em might} take, as in common constraint
systems; i.e., they do not state $D$, but negated $D$), and (4)
success information: {\tt yes} and {\tt no} stand for {\em
success} and {\em failure}, respectively. Finally, `{\tt ;}'
separates the solutions which has been explicitly requested by the
user. Primitive constraints in the finite domain constraint store
are not shown.
\end{sloppypar}

\end{example}

\begin{example}
We show a $\Toyfd$ program involving labeling to solve the
classical N-queens problem whose objective is to place $N$ queens
on an $N \times N$ chessboard so that there are no threatening
queens.

\begin{verbatim}
include "misc.toy"
include "cflpfd.toy"

queens :: int -> [labelType] -> [int]
queens N Label = L <== length L == N, domain L 1 N,
                       constrain_all L, labeling Label L

constrain_all :: [int] ->  bool
constrain_all [] = true
constrain_all [X|Xs] = true <== constrain_between X Xs 1,
                       constrain_all Xs

constrain_between :: int -> [int] -> int -> bool
constrain_between X [] N = true
constrain_between X [Y|Ys] N = true <== no_threat X Y N,
                       constrain_between X Ys (N+1)

no_threat:: int -> int -> int -> bool
no_threat X Y I = true <== X #\= Y, X #+ I #\= Y, X #- I #\= Y
\end{verbatim}

The intended meaning of the functions should be clear from their
names and definitions, provided that {\tt length L} returns the
length of the list {\tt L}. The first two lines are needed to
include predefined functions such as {\tt length} and {\tt domain}. An
example of solving at the command prompt, where {\tt ff} stands for
the first-fail enumeration strategy (see Section~\ref{labelings}), is
\begin{verbatim}
TOY(FD)> queens 15 [ff] == L
         yes    L ==  [1,3,5,14,11,4,10,7,13,15,2,8,6,9,12]
\end{verbatim}
\end{example}

\begin{example}
We present a $\Toyfd$ program using syntactic sugaring for
predicate-like functions that solves the well-known $\Clpfd$
program {\em Send+More=Money}.

\begin{verbatim}
smm :: int -> int -> int -> int -> int -> int -> int -> int
       -> [labelType] -> bool

smm S E N D M O R Y Label :-
       domain [S,E,N,D,M,O,R,Y] 0 9, S #> 0, M #> 0,
       all_different [S,E,N,D,M,O,R,Y], add S E N D M O R Y,
       labeling Label [S,E,N,D,M,O,R,Y]

add :: int -> int -> int -> int -> int -> int -> int -> int -> bool
add S E N D M O R Y :-  1000#*S #+ 100#*E #+ 10#*N #+ D
                     #+ 1000#*M #+ 100#*O #+ 10#*R #+ E
        #=  10000#*M #+ 1000#*O #+ 100#*N #+ 10#*E #+ Y
\end{verbatim}
\end{example}

For our simple $\Toyfd$ programs, some examples of goals and
answers which can be computed by $\Toyfd$ are shown in
Table~\ref{table:Solving goals in Toy}.

\begin{table}[htbp]
\caption{Examples of Goal Solving} \label{table:Solving goals in Toy}
\begin{center}
\begin{tabular}{l}
      \hline\hline
      \begin{tabular}{ll}
        \hspace{1cm} {\em Goal} \hspace{5cm} &{\em Answers}\\
      \end{tabular}\\
      \hline
      \begin{tabular}{l|l}
               {\tt domain [A,B] 1 (1+2),~A\#>B,} & {\tt A==2,B==1; A==3,B==1;}  \\
               {\tt all\_different [A,B], labeling [~] [A,B]} & {\tt A==3,B==2; no more solutions}\\
    \hline
               {\tt domain [X,Y,Z] 1 10,} & {\tt X in 1..2,Y==1,Z in 8..10;}\\
               {\tt 2 \#* X \#+ 3 \#* Y \#+ 2 \#< Z} & {\tt no more solutions}\\
    \hline
               {\tt domain [X,Y,Z] 1 5, X \#> Y,} & {\tt X in 4..5,Y in 3..4,Z in 1..3;} \\
               {\tt 2 \#* Y \#> Z \#+ 4, X \#>= Z} & {\tt no more solutions}\\
    \hline
               {\tt smm S E N D M O R Y [] == T} & {\tt S==9,E==5,N==6,D==7,M==1,O==0,}\\
                                                 & {\tt R==8,Y==2,T==true;}   \\
                                                 & {\tt no more solutions}\\
    \hline
               {\tt queens 5 [] == [M,A,E,Y,B],} & {\tt M==1,A==3,E==5,Y==2,B==4,S==9,} \\
               {\tt smm S E N D M O R Y []}      & {\tt N==6,D==7,O==0,R==8;}\\
                                                 & {\tt no more solutions}\\
      \end{tabular}\\
      \hline\hline
    \end{tabular}
\end{center}
\end{table}
\subsection{$\Cflpfd$ vs. $\clpfd$} \label{sect:user level advantages}

It is commonly acknowledged that $\Clpfd$ is a successful
declarative approach; hence, we discuss the advantages of
$\Cflpfd$, focusing on the $\Toyfd$ implementation, with respect
to $\clpfd$. This section explains why the addition of $FP$
features enhances the $CLP$ setting. When necessary, we illustrate
different features of $\Cflpfd$ by means of examples. Further
programming examples in pure functional logic programming and
$\Cflpfd$ can be found, respectively, in
\cite{fraguas+:multiparadigm-system-rta99}
and~\cite{fernandez+:toy(fd)-manual02}.

\subsubsection{$\Cflpfd$ $\supset$ $\clpfd$}

As already pointed out, besides other features, $\Cflpfd$ provides
the main characteristics of $\clpfd$, i.e., ${\cal FD}$ constraint
solving, non-determinism and relational form. Moreover, $\Cflpfd$
provides a sugaring syntax for $LP$ predicates and thus, as
already commented, any pure $\clpfd$-program can be
straightforwardly translated into a $\cflpfd$-program. In this
sense, $\Clpfd$ may be considered as a strict subset of $\cflpfd$
with respect to problem formulation. As a direct consequence, our
language is able to cope with a wide range of applications (at
least with all those applications that can be formulated with a
$\Clpfd$ language). We will not insist here on this matter, but
prefer to concentrate on the extra capabilities of $\Cflpfd$ with
respect to $\clpfd$.

\subsubsection{$\cflpfd$ ~$\setminus$ ~$\Clpfd$}

\begin{sloppypar}
Due to its functional component, $\Cflpfd$ adds further
expressiveness to $\Clpfd$ as allows the declaration of functions
and their evaluation in the $FP$ style. In the following, we
enumerate and discuss other features not present (or unusual) in
the $\Clpfd$ paradigm.
\end{sloppypar}

\paragraph{\bf Types.}

Our language is strongly typed and thus involves all the
well-known advantages of a type checking process, enhancing
program development and maintenance. Each ${\cal FD}$ constraint
has associated, like any function, a type declaration, which means
that a wrong use can be straightforwardly detected in the typical
type checking process.

\begin{sloppypar}
\paragraph{\bf Functional Notation.}

It is well-known that functional notation reduces the number of
variables with respect to relational notation, and thus, $\Cflpfd$
increases the expressiveness of $\Clpfd$ as it combines relational
and functional notation. For instance, in $\Clpfd$ the constraint
conjunction  {\tt N=2, X $\in$ [1,10-N]} cannot be expressed
directly and must be written adding a third component, as {\tt
N=2, Max is 10-N, domain([X],1,Max)} that uses an extra variable.
However, $\Toyfd$ expresses that constraint directly as {\tt N==2,
domain [X] 1 (10-N)}.
\end{sloppypar}

\paragraph{\bf Currying.}

Again, due to its functional component, $\Toyfd$ allows curried
functions (and thus constraints); for instance, see the application
of curried ${\cal FD}$ constraint {\tt (3\#$<)/1$} in
Example~\ref{ex: map} later in this section.

\paragraph{\bf Higher-Order and Polymorphism.}

In $\Toyfd$ functions are first-class citizens, which means that a
function (and thus an ${\cal FD}$ constraint) can appear in  any
place where data do. As a direct consequence, an ${\cal FD}$
constraint may appear as an argument (or even as a result) of
another function or constraint. The functions managing other
functions are called higher-order (HO) functions. Also,
polymorphic arguments are allowed in $\cflpfd$.

\begin{example}\label{ex: map}
A traditional example of a polymorphic HO function is
\begin{verbatim}
map :: (A -> B) -> [A] -> [B]
map F [] = []
map F [X|Xs] = [(F X) | (map F Xs)]
\end{verbatim}
that receives both a function F and a list as arguments and
produces a list resulting from applying the function to each
element in the list. Now, suppose that $X$ and $Y$ are ${\cal FD}$
variables ranging in the domain [0..100] (expressed, for instance,
via the constraint {\tt domain [X,Y] 0 100}). Then, the goal {\tt
map (3\#$<$) [X,Y]} returns the Boolean list {\tt [true,true]}
resulting from evaluating the list {\tt [3\#<X,3\#<Y]}, and {\tt
X} and {\tt Y} are also restricted to have values in the range
{\tt [4,100]} as the constraints ~{\tt 3\#<X}~ and ~{\tt 3\#<Y}~
are sent to the constraint solver. Note also the use of the
curried function {\tt (3\#<)}.
\end{example}

\paragraph{\bf Laziness.} In contrast to logic languages, functional languages
support {\em lazy evaluation}, where function arguments are
evaluated to the required extend (the {\em call-by-value} used in
$LP$ vs. the {call-by-need} used in $FP$). Strictly speaking, lazy
evaluation may also correspond to the notion of {\em only once
evaluated} in addition to {\em only required extent}
\cite{peytonjones:implementation-functional-programming-prectice87}.
$\Toyfd$ increases the power of $\Clpfd$ by incorporating a novel
mechanism that combines {\em lazy evaluation} and ${\cal FD}$
constraint solving, in such a way that only the demanded
constraints are sent to the solver. This is a powerful mechanism
that opens new possibilities for ${\cal FD}$ constraint solving.
For example, in contrast to $\clpfd$, it is possible to manage
infinite structures.

\begin{example}
\label{ex: generateFD}
Consider the recursive functions {\tt take} and {\tt generateFD} from Example \ref{lazy}.
An eager evaluation of the following goal does not terminate as it tries to
completely evaluate the second argument, yielding to an infinite
computation. However, a lazy evaluation generates just the first 3
elements of the list, as shown below:
\begin{verbatim}
TOY(FD)> take 3 (generateFD 10) == List
         yes    List ==  [ _A, _B, _C ]   _A, _B, _C in 0..9
\end{verbatim}
\end{example}

In general, lazy narrowing avoids computations which are not
demanded, therefore saving computation time. Example~\ref{ex: magic
series} contains a formulation of the typical magic series (or
sequences)  problem \cite{hentenryck:cs-in-lp-mit89}. This example
highlights the expressive power of $\Toyfd$ by solving multiple
problem instances that can be described and solved via lazy
evaluation of infinite lists.

\begin{example}\label{ex: magic series}
Let {\tt S = (s$_0$, s$_1$, \ldots,s$_{N-1}$)} be a non-empty
finite series of non-negative integers. The series {\tt S}  is
said {\em N-magic} if and only if there are {\tt s$_i$}
occurrences of {\tt i} in {\tt S}, for all {\tt i $\in$
\{0,\ldots,N-1\}}. Below, we propose a $\Toyfd$ program to
calculate magic series where the function {\tt generateFD} is as
defined in Example~\ref{lazy}.

\begin{verbatim}
lazymagic :: int -> [int]
lazymagic N = L <== take N (generateFD  N) == L,
                    constrain L  L  0  Cs, sum L (#=) N,
                    scalar_product Cs  L (#=) N, labeling [ff] L

constrain :: [int] -> [int] -> int -> [int] -> bool
constrain [] A B  [] = true
constrain [X|Xs] L I [J|Js] = true <== I==J, count I L (#=) X,
                                       constrain Xs L (I+1) Js
\end{verbatim}
{\tt sum/3}, {\tt scalar\_product}/4 and {\tt count}/4 are
predefined HO constraints \cite{fernandez+:toy(fd)-manual02}, that
accept a relational ${\cal FD}$ constraint operator with type {\tt
int} $\rightarrow$ {\tt int} $\rightarrow$ {\tt bool} as argument
(e.g., the constraint {\tt \#=}). {\tt sum L C N} means that the
summation of the elements in the list {\tt L} is related through
{\tt C} with the integer {\tt N} (in the example, the summation is
constrained to be equal to {\tt N}). {\tt scalar\_product} and {\tt
count} stand for scalar product and element counting under the same
parameters as {\tt sum}.

A goal {\tt lazymagic N}, for some natural N, returns the N-magic
series where the condition {\tt take N (generateFD N)} is
evaluated lazily as {\tt (generateFD N)} produces an infinite
list. More interesting is to return a list of different solutions
starting from N. This can be done using a recursive definition to
produce the infinite list of magic series {\tt (from N)} as shown
below.

\begin{verbatim}
magicfrom :: int -> [[int]]
magicfrom N = [lazymagic N | magicfrom (N+1)]
\end{verbatim}

Now, it is easy to generate a list of magic series by lazy
evaluation. For example, the following goal generates a 3-element
list containing, respectively, the solution to the problems of
7-magic, 8-magic and 9-magic series.

\begin{verbatim}
TOY(FD)> take 3 (magicfrom 7) == L
          yes      L ==  [ [ 3, 2, 1, 1, 0, 0, 0 ],
                           [ 4, 2, 1, 0, 1, 0, 0, 0 ],
                           [ 5, 2, 1, 0, 0, 1, 0, 0, 0 ] ]
\end{verbatim}

More expressiveness is shown by mixing curried functions, HO
functions and function composition (another nice feature from the
functional component of $\toyfd$). For example, consider the
$\Toyfd$ code shown below:

\begin{verbatim}
from :: int -> [int]
from N = [N | from (N+1)]

(.):: (B -> C) -> (A -> B) -> (A -> C)
(F . G) X = F (G X)

lazyseries :: int -> [[int]]
lazyseries = map lazymagic . from
\end{verbatim}
where {\tt (.)}/2 defines the composition of functions. Observe
that {\sl lazyseries} curries the composition {\tt (map
lazymagic) . from}. Then, it is easy to generate the 3-element list
shown above by just typing the goal
\begin{verbatim}
TOY(FD)> take 3 (lazyseries 7) == L
\end{verbatim}
\end{example}

This simple example gives an idea of the nice features of
$\Cflpfd$ that combines ${\cal FD}$ constraint solving, management
of infinite lists and lazy evaluation, curried notation of
functions, polymorphism, HO functions (and thus HO constraints),
composition of functions and a number of other characteristics
that increase the potentialities with respect to $\clpfd$.

\subsection{Correctness of the \Cflpfd Implementation}

In this section, we briefly discuss the correctness of our
$\Toyfd$ implementation with respect to our $\cflpfd$ framework.

$\Toyfd$ integrates, as a host language, the higher-order lazy
functional logic language $\Toy$
\cite{fraguas+:multiparadigm-system-rta99} and, as constraint
solver, the efficient ${\cal FD}$ constraint solver of SICStus
Prolog \cite{carlsson+:open-ended-plilp97}. Under the condition of
considering just an empty set of protected variables, the SICStus
Prolog finite domain solver always satisfies the conditions for
constraint solvers required in Section~\ref{solvers}. Since the
$CLNC(\cdom)$ calculus is {\em strongly complete} (see
\cite{fraguas+:lazy-narrowing-calculus-for-declarative-cp-ppdp04})
in the sense that the choice of goal transformation rules can be a
{\em don't care} choice, in practice, we can choose a suitable
demand-driven strategy: our $\cdom$ constraint solver is only
applied at the end of the process of goal solving, when we have an
empty set of protected variables (as we have done in the example
in Section $3.3$) or when protected variables are not relevant.
This strategy can be performed in the $CLNC(\cdom)$ calculus in
the line of \cite{vado+:frocos05} as well as in $\toyfd$ in the
line of \cite{soniadelvado+:wcflp05}. Therefore, we can conclude
that our operational semantics with this strategy covers
adequately the $\Toyfd$ implementation.

Additionally, we have run a number of tests in the implementation
and have compared the derivations produced by the calculus
$CLNC(\cdom)$ to the traces obtained from debugging in $\toyfd$,
and the results show that these are effectively identical by
following an adequate demand-driven strategy in $CLNC(\cdom)$. For
instance, the $\Cflpfd$ program shown in Example~\ref{lazy}
corresponds almost directly to a $\Toyfd$ program, and the solving
of the goal $check\_list\ (from\ M)\ <\ 3$ in $\Toyfd$ is shown
below (see \cite{soniadelvado+:wcflp05} for more details).

\begin{verbatim}
Toy(FD)> check_list (from M) < 3
      yes
      M in 1..2
      Elapsed time: 0 ms.

more solutions (y/n/d) [y]?
      yes
      M in 3..4
      Elapsed time: 0 ms.

more solutions (y/n/d) [y]?
      no.
      Elapsed time: 0 ms.
\end{verbatim}

\noindent Note that the computed answers correspond exactly to
those obtained in the goal solving process described in
Section~\ref{example of goal resolution} via the narrowing
calculus $CLNC(\cdom)$.

\subsection{Notes about the Implementation}
\label{sect:details about implementation}

\nc{\com}{\diamondsuit}

In $\toyfd$, ${\cal FD}$ constraints are evaluated internally by
using mainly two predicates: {\tt hnf(E,H)}, which specifies that
{\tt H} is one of the possible results of narrowing the expression
{\tt E} into head normal form, and {\tt solve}/1, which checks the
satisfiability of constraints (of rules and goals) before the
evaluation of a given rule. This predicate is, basically, defined as
follows\footnote{The code does not correspond exactly to the
implementation, which is the result of many transformations and
optimizations.}:

\begin{quote}
$\begin{array}{lll} (1)\ \mathtt{solve((\varphi, \varphi'))} &
\mbox{:$-$}
                 & \mathtt{solve(\varphi), solve(\varphi').} \\
(2)\ \mathtt{solve( L == R )}            & \mbox{:$-$}
                 &  \mathtt{hnf(L, L'), hnf(R, R'), equal(L', R').} \\
(3)\ \mathtt{solve( L ~/= R )}           & \mbox{:$-$}
                 & \mathtt{hnf(L, L'), hnf(R, R'), notequal(L', R').} \\
(4)\ \mathtt{solve( L \#\com ~ R )}      & \mbox{:$-$}
                 & \mathtt{hnf(L, L'), hnf(R, R'), \{L' \#\com R'\}.}  \\
                                         &
                 & \qquad \text{where } \mathtt{\com \in \{=,
\backslash\!\!=, <, <\!=, >, >\!=\}.} \\ (5)\ \mathtt{solve( C ~ A_1
\ldots A_n )} & \mbox{:$-$}
                 & \mathtt{hnf(A_1, A_1'),\ldots,hnf(A_n,
A_n'),\{C(A_1',\ldots,A_n')\}.}    \\
                                          &
                 & \qquad \text{where {\tt C} is any constraint returning a Boolean. }
\end{array}$
\end{quote}

\normalsize  The interaction with the constraint solver (i.e.,
SICStus ${\cal FD}$ constraint solver in the current $\Toyfd$
version) is reflected in the last two clauses: every time an ${\cal
FD}$ constraint appears, the solver is eventually invoked with a
goal $\{G\}$ where $G$ is the translation of the ${\cal FD}$
constraint from $\Toyfd$ to SICStus Prolog. Head normal forms are
required for constraint arguments in order to allow the solver to
solve the constraint.

\subsection{Performance}\label{sect: Efficiency Compared}

As far as we know, $\Toyfd$ was the first $FLP$ system integrating
a ${\cal FD}$ constraint system. However, we know about the
existence of an implementation of the $FLP$ language
Curry~\cite{hanus:curry99} that supports a limited set of ${\cal
FD}$ constraints~\cite{hanus:pakcs-manual05}. This implementation,
called PAKCS, provides the following constraints:

(1) a set of arithmetical operations {\tt
\{*\#,+\#,-\#,=\#,/=\#,<\#,<=\#,>\#,>=\#\}},

(2) a membership constraint {\tt domain} /3,

(3) some global constraints\footnote{Exactly, those named in this
paper, i.e.  {\tt all\_different}/1,  {\tt count}/4, {\tt
scalar\_product}/4 and {\tt sum}/3.} and

(4) an enumeration constraint {\tt labeling} /1 that also provides
searching options.

In this section, we compare the performance of $\Toyfd$ with that
of the {\em Curry2Prolog} compiler, which is the most efficient
implementation of Curry inside PAKCS (version 1.7.1 of December
2005).

In addition, for evaluating if $\Toyfd$ is competitive with
respect to existing $\clpfd$ systems, we have also considered four
well-known $\clpfd$ systems:
\begin{enumerate}
\item The version 3.12.1 of April 2005 of the ${\cal FD}$ constraint
solver of SICStus Prolog
\cite{carlsson+:open-ended-plilp97,sicstus-homepage}. This solver
was included in order to measure the overhead due to the
management of functional logic expressions, which are compiled to
SICStus Prolog in $\toyfd$, and, therefore, including all the
stuff needed to handle the $FLP$ characteristics such as laziness
and higher-order functions.

\item The GNU Prolog system (version 1.2.16) \cite{diaz+:GNU-Prolog-jflp01,gnu-prolog-homepage}, which
is a free Prolog compiler that includes
one of the most efficient finite domain constraint solver. This solver is based on the concept of {\em indexicals} \cite{codognet+:clp(fd)-jlp96}
and it has been demonstrated that it has a performance comparable to commercial systems.

\item SWI-Prolog (version 5.4.x) \cite{Wielemaker:swi-prolog-wlpe03,swi-prolog-homepage} that it is
   an emergent and very promising Prolog system that provides
   an integer domain constraint solver implemented with attributed variables.

\item Ciao Prolog (version 1.10\#5 of August 2004) which is a full multi-paradigm programming environment
for developing programs in the Prolog language and in several
other languages which are extensions and modifications of Prolog
in several interesting and useful directions. Ciao Prolog provides
a package, based upon the indexical concept, to write and evaluate
constraint programming expressions over finite domains in a Ciao
program.

\end{enumerate}

\subsubsection{Labeling} \label{labelings}

Constraint solving can be implemented with a combination of two
processes: constraint propagation and labeling (i.e., search)
\cite{dechter:constraint-processing-morgan-kauffmann-2003}. The
labeling process consists of (1) choosing a variable (variable
ordering) and (2) assigning to the variable  a value which belongs
to its domain (value ordering). The variable ordering and the
value ordering used for the labeling can considerably influence
the efficiency of the constraint solving when only one solution to
the problem is required. It has little effect when the search is
for all solutions. In this study, we consider two labelings, the
na\"{\i}ve labeling that chooses the leftmost variable of a list
of variables and then selects the smallest value in its domain,
and the {\em first-fail} labeling that uses a principle
\cite{haralick+:increasing-ff-ai80} which says that {\em to
succeed, try first where you are the most likely to fail}. This
principle recommends the choice of the most constrained variable,
which often means (for the finite domain) choosing a variable with
the smallest domain. The na\"{\i}ve labeling assures that both
variable and value ordering are the same for all the systems and
hence (although less efficient) is better for comparing the
different systems when only one solution is required.

\subsubsection{The Benchmarks}

We have used a wide set of benchmarks\footnote{All the programs
used in the comparison are available at {\em
http://www.lcc.uma.es/$\sim$afdez/cflpfd/}.} and, for the sake of
fairness, whenever it was possible, we used exactly the same
formulation of the problems for all systems as well as the same
${\cal FD}$ constraints. The benchmarks used are:

\begin{itemize}
    \item {\bf cars}: solve a car sequencing problem with 10 cars \cite{dincbas+:car-sequencing-clp-ecai88}. This
benchmark deals  with 100 Boolean variables (i.e., finite domain
variables  ranging over [0,1]), 10 finite domain variables ranging
over [1,6], 6 {\em atmost} constraints, 50 {\em element}
constraints, and 49 linear disequations.

    \item {\bf equation 10}: a system of 10 linear equations with 7
variables ranging over [0,10].

    \item {\bf equation 20}: a system of 20 linear equations with 7
variables ranging over [0,10].

    \item {\bf magic series (N)}: calculate a series of $N$ numbers
such that each of them is the number of occurrences in the series
of its position in the series \cite{codognet+:clp(fd)-jlp96}.

    \item {\bf optimal Golomb ruler (N)}: find an ordered set of $n$ distinct
non-negative integers, called \emph{marks}, $a_1 < ... < a_n$,
such that all the differences $a_i-a_j$ ($i > j$) are distinct and
$a_n$ is minimum  \cite{shearer:new-optimium-golomb-rulers-ieee-tit90}.

    \item {\bf queens (N)}: place $N$ queens on a $N\times N$
chessboard such that no queen attacks each other \cite{hentenryck:cs-in-lp-mit89}.

    \item {\bf pythagoras}: calculate the proportions of a triangle by using
the Pythagorean theorem. This problem involves 3 variables ranging
over [1,1000], and 7 disequality (non-linear) equations.

    \item {\bf sendmore}: a cryptoarithmethic problem with 8
variables ranging over [0,9], with one linear equation, 2
disequations and 28 inequality constraints (or alternatively one
{\em all\_different} constraint imposed over the whole set of
constrained variables). It consists of solving the equation $SEND
+ MORE = MONEY$.

       \item {\bf suudoku}: the problem is to fill partially filled 9x9 squares of 81 squares such that
           each row and column are permutations of [1,...,9], and each 3x3
           square, where the leftmost column modulo 3 is 0, is a permutation
           of [1,...,9].
\end{itemize}

The programs {\bf equation 10}, {\bf equation 20} and {\bf
sendmore} test the efficiency of the systems to solve linear
equation problems. The programs {\bf cars} and {\bf suudoku} check
the efficiency of specialized constraints such as the {\em
all\_different} constraint. The {\bf pythagoras} problem deals
with non-linear equations.

The {\bf queens} and {\bf magic series} programs are scalable and
therefore useful to test how the systems work for bigger instances
of the same problem. Note that both the number of variables and
the number of values for each variable grow linearly with the parameter $N$ in the examples.
That is, given a value $N$, at least $N$ ${\cal FD}$ variables
must be declared with domains that range between 0 or 1 and $N$.

The search for {\bf optimal Golomb rulers} is an extremely
difficult task as it is a combinatorial problem whose bounds grow
geometrically with respect to the solution size
\cite{shearer:new-optimium-golomb-rulers-ieee-tit90}. This (also
scalable) benchmark allows us to check the optimization
capabilities of the system.

\begin{table}[htb]
\caption{$\Toyfd$ vs. C(F)LP Systems:  Na\"{\i}ve Labeling}
\centerline{
\begin{tabular}{lrrrrrr} \hline\hline
   {\bf Benchmark}      & {\em $\toyfd$}  & {PAKCS}      & {\em SICStus}       &  SWI                      &  GNU                  & Ciao      \\ \hline
   {\bf cars}           & 5               & N            & 5                   & N                         & 1                     & N     \\
   {\bf equation10}     & 20              & 50           & 10                  & 590                       & 2                     & -     \\
   {\bf equation20}     & 35              & 60           & 10                  & 1185                      & 4                     & -     \\
   {\bf magic(64)}      & 265             & 340          & (430) 260           & N (OGS)                   & 134                   & N     \\
   {\bf magic(100)}     & 910             & 980          & (1520) 900          & N (OGS)                   & 901                   & N     \\
   {\bf magic(150)}     & 2700            & 3180         & (4770) 2560         & N (OGS)                   & (SO) 4894             & N     \\
   {\bf magic(200)}     & 5970            & 6540         & (10870) 5690        & N (OGS)                   & (SO) 14570            & N     \\
   {\bf magic(300)}     & 18365           & 22750        & (RE) 17780          & N (OGS)                   & (SO) 68020            & N     \\
   {\bf pythagoras}     & 50              & 80           & 20                  & 940                       & 10                    & 902     \\
   {\bf queens(8)}      & 10              & 20           & 10                  & 110                       & 1                     & 31     \\
   {\bf queens(16)}     & 180             & 200          & 170                 & 38720                     & 11                    & 6873     \\
   {\bf queens(20)}     & 4030            & 4200         & 3930                & 1064130                   & 216                   & 190435     \\
   {\bf queens(24)}     & 8330            & 8400         & 8120                & ??                        & 460                   & 576625     \\
   {\bf queens(30)}     & 1141760         & 1141940      & 1069750             & ??                        & 67745                 & ??     \\
   {\bf sendmore}       & 0               & 5            & 0                   &  15                       & 0                     & 14     \\
   {\bf suudoku}        & 10              & 20           & 10                  &  60                       & 1                     & 51     \\\hline\hline
\end{tabular}}\label{naive labeling results}
\end{table}

\begin{table}[htb]
\caption{Speed-Up of $\Toyfd$ wrt. other C(F)LP Systems for
Na\"{\i}ve Labeling} \centerline{
\begin{tabular}{lrrrrr}\hline\hline
   {\bf Benchmark}      & PAKCS                           & {\em SICStus}                    & SWI                      & GNU                     &  Ciao \\ \hline
   {\bf cars}           & $\infty$                       & 1 00                              & $\infty$                 & 0.20                    &  $\infty$  \\
   {\bf equation10}     & 2.50                           & 0.50                              & 29.50                    & 0.10                    &  $\infty$  \\
   {\bf equation20}     & 1.71                           & 0.28                              & 33.85                    & 0.11                    &  $\infty$  \\
   {\bf magic (64)}     & 1.28                           & (1.62) 0.98                       & $\infty$                 & 0.50                    &  $\infty$  \\
   {\bf magic (100)}    & 1.07                           & (1.67) 0.98                       & $\infty$                 & 0.99                    &  $\infty$  \\
   {\bf magic (150)}    & 1.17                           & (1.76) 0.98                       & $\infty$                 & ($\infty$) 1.81         &  $\infty$   \\
   {\bf magic (200)}    & 1.09                           & (1.82) 0.99                       & $\infty$                 & ($\infty$) 2.44         &  $\infty$   \\
   {\bf magic (300)}    & 1.23                           & ($\infty$) 0.96                   & $\infty$                 & ($\infty$) 3.70         &  $\infty$   \\
   {\bf pythagoras}     & 1.60                           & 0.40                              & 18.80                    & 0.20                    &  18.04 \\
   {\bf queens (8)}     & 2.00                           & 1.00                              & 11.00                    & 0.10                    &  3.12  \\
   {\bf queens (16)}    & 1.11                           & 0.94                              & 215.11                   & 0.06                    &  38.18 \\
   {\bf queens (20)}    & 1.04                           & 0.97                              & 264.05                   & 0.05                    &  42.25 \\
   {\bf queens (24)}    & 1.00                           & 0.97                              & (?)                      & 0.05                    &  69.22 \\
   {\bf queens (30)}    & 1.00                           & 0.93                              & (?)                      & 0.05                    &  (?)  \\
   {\bf sendmore}       & $\geq$ 5.00                    & $\geq$ 1.00                       & $\geq$ 15.00             & $\geq$ 1.00             &  $\geq$ 14.00 \\
   {\bf suudoku}        & 2.00                           & 1.00                              & 6.00                     & 0.10                    &  5.10  \\\hline\hline
\end{tabular}}\label{naive labeling normalized results}
\end{table}

\begin{table}[htb]
\caption{$\Toyfd$ vs. C(F)LP systems: First-Fail Labeling}
\centerline{
\begin{tabular}{lrrrrrr} \hline\hline
   {\bf Benchmark}      & {\em $\toyfd$}   & PAKCS  & {\em SICStus}   & SWI        &  GNU      & Ciao    \\ \hline
   {\bf cars}           & 0                & N      & 0               & N          & 0         & N       \\
   {\bf equation10}     & 20               & 50     & 10              & N          & 3         & N      \\
   {\bf equation20}     & 30               & 55     & 15              & N          & 4         & N      \\
   {\bf magic (64)}     & 90               & 150    & (320) 80        & N          & 18        & N      \\
   {\bf magic (100)}    & 220              & 310    & (1090) 195      & N          & 53        & N      \\
   {\bf magic (150)}    & 470              & 690    & (3440) 465      & N          & (SO) 52   & N      \\
   {\bf magic (200)}    & 870              & 1480   & (7950) 850      & N          & (SO) 125  & N      \\
   {\bf magic (300)}    & 1835             & 3610   & (RE) 1820       & N          & (SO) 568  & N      \\
   {\bf magic (400)}    & 3420             & 10050  & (RE) 3370       & N          & (SO) 1088 & N      \\
   {\bf magic (500)}    & 5510             & 13100  & (RE) 5250       & N          & (SO) 1830 & N      \\
   {\bf pythagoras}     & 50               & 80     & 10              & N          & 10        & N      \\
   {\bf queens (8)}     & 10               & 15     & 5               & N          & 1         & N      \\
   {\bf queens (16)}    & 20               & 50     & 8               & N          & 2         & N      \\
   {\bf queens (20)}    & 45               & 75     & 10              & N          & 3         & N      \\
   {\bf queens (24)}    & 40               & 80     & 15              & N          & 4         & N      \\
   {\bf queens (30)}    & 150              & 190    & 25              & N          & 6         & N      \\
   {\bf sendmore}       & 0                & 5      & 0               & N          & 0         & N      \\
   {\bf suudoku}        & 10               & 20     & 10              & N          & 1         & N      \\\hline\hline
\end{tabular}}\label{first fail labeling results}
\end{table}

\begin{table}[htb]
\caption{Speed-Up of $\Toyfd$ wrt. other C(F)LP Systems for
First-Fail Labeling} \centerline{
\begin{tabular}{lrrrrrr} \hline\hline
   {\bf Benchmark}      & PAKCS                   & {\em SICStus}            & SWI                     & GNU                   &  Ciao                   & {\em $\toyfd$} \\
                        &                         &                          &                         &                       &                         & (na\"{\i}ve) \\ \hline
   {\bf cars}           & $\infty$                &  $\geq$ 1.00             & $\infty$                &  $\geq$ 1.00          & $\infty$                &    $\geq$ 5.00                                    \\
   {\bf equation10}     & 2.50                    &  0.50                    & $\infty$                &  0.15                 & $\infty$                &    1.00                                           \\
   {\bf equation20}     & 1.83                    &  0.50                    & $\infty$                &  0.13                 & $\infty$                &    1.16                                           \\
   {\bf magic (64)}     & 1.66                    &  (3.55) 0.88             & $\infty$                &  0.20                 & $\infty$                &    2.94                                           \\
   {\bf magic (100)}    & 1.40                    &  (4.95) 0.88             & $\infty$                &  0.24                 & $\infty$                &    4.13                                           \\
   {\bf magic (150)}    & 1.46                    &  (7.31) 0.98             & $\infty$                &  ($\infty$) 0.11      & $\infty$                &    5.74                                           \\
   {\bf magic (200)}    & 1.70                    &  (9.13) 0.97             & $\infty$                &  ($\infty$) 0.14      & $\infty$                &    6.86                                           \\
   {\bf magic (300)}    & 1.96                    &  ($\infty$) 0.99         & $\infty$                &  ($\infty$) 0.30      & $\infty$                &    10.00                                          \\
   {\bf magic (400)}    & 2.93                    &  ($\infty$) 0.98         & $\infty$                &  ($\infty$) 0.31      & $\infty$                &    (?)                                             \\
   {\bf magic (500)}    & 2.37                    &  ($\infty$) 0.95         & $\infty$                &  ($\infty$) 0.33      & $\infty$                &    (?)                                             \\
   {\bf pythagoras}     & 1.60                    &  0.20                    & $\infty$                &  0.20                 & $\infty$                &    1.00                                           \\
   {\bf queens (8)}     & 1.50                    &  0.50                    & $\infty$                &  0.10                 & $\infty$                &    1.00                                           \\
   {\bf queens (16)}    & 2.50                    &  0.40                    & $\infty$                &  0.10                 & $\infty$                &    9.00                                           \\
   {\bf queens (20)}    & 1.66                    &  0.22                    & $\infty$                &  0.06                 & $\infty$                &    89.55                                          \\
   {\bf queens (24)}    & 2.00                    &  0.37                    & $\infty$                &  0.10                 & $\infty$                &    208.25                                         \\
   {\bf queens (30)}    & 1.26                    &  0.16                    & $\infty$                &  0.04                 & $\infty$                &    7611.73                                        \\
   {\bf sendmore}       & $\geq$ 5.0              &  $\geq$ 1.00             & $\infty$                &  $\geq$ 1.00          & $\infty$                &    $\geq$ 1.00                                    \\
   {\bf suudoku}        & 2.00                    &  1.00                    & $\infty$                &  0.10                 & $\infty$                &    1.00                                           \\\hline\hline
\end{tabular}}\label{normalized results for first fail labeling}
\end{table}

\begin{table}[htb]
\caption{$\Toyfd$ vs. C(F)LP Systems: Optimization Benchmarks}
\centerline{
\begin{tabular}{lrrrrrr} \hline\hline
   {\bf Benchmark}      & {\em $\toyfd$}  & {PAKCS}      & {\em SICStus}       &  SWI                      &  GNU                & Ciao         \\ \hline
   {\bf golomb(8)}      & 360             & 350          & 280                 & N                         & 86                  & N       \\
   {\bf golomb(10)}     & 26230           & 27500        & 25730               & N                         & 8595                & N       \\
   {\bf golomb(12)}     & 5280170         & 5453220      & 5208760             & N                         & 2162863             & N       \\\hline\hline
\end{tabular}}\label{optimization results}
\end{table}

\begin{table}[htb]
\caption{Speed-Up of $\Toyfd$ wrt. other C(F)LP Systems for
Optimization Benchmarks} \centerline{
\begin{tabular}{lrrrrr} \hline\hline
   {\bf Benchmark}      & {\em SICStus}             & PAKCS                         & SWI                         & GNU                    &  Ciao \\ \hline
   {\bf golomb(8)}      & 0.77                      & 0.97                          & $\infty$                    & 0.23                   &  $\infty$   \\
   {\bf golomb(10)}     & 0.98                      & 1.04                          & $\infty$                    & 0.32                   &  $\infty$   \\
   {\bf golomb(12)}     & 0.98                      & 1.03                          & $\infty$                    & 0.40                   &  $\infty$   \\\hline\hline
\end{tabular}}\label{optimization results + normalization}
\end{table}

\subsubsection{Results}

All the benchmarks were performed on the same Linux machine (under
Fedora Core system, 2.69-1667) with an Intel(R) Pentium 4
processor running at 2.40 GHz and with a RAM memory of 512 Mb. For
the sake of brevity, we only provide the results for first
solution search.

Table~\ref{naive labeling results} shows the results using
na\"{\i}ve labeling. The meaning for the columns is as follows.
The first column gives the name of the benchmark used in the
comparison, and the next six columns show the running (elapsed)
time (measured in milliseconds) to find the first answer of the
benchmark for each system.

Table~\ref{naive labeling normalized results} shows the results
shown in Table~\ref{naive labeling results} in terms of the
speed-up of $\toyfd$ with respect to the rest of the systems (that
is, the result of dividing the time of a given system by the time
of $\toyfd$).

Table~\ref{first fail labeling results} shows the results of
solving the same benchmarks by using first-fail labeling. Note
that the current versions of SWI Prolog and Ciao Prolog do not
provide first-fail labeling. Also, Table~\ref{normalized results
for first fail labeling} shows the speed-up corresponding to the
results in Table~\ref{first fail labeling results} and again
displays the performance of $\toyfd$ with respect to the rest of
the systems. The meaning for the columns is as in Table~\ref{naive
labeling normalized results}, but a last column is added in order
to show the speed-up of $\Toyfd$ using first-fail labeling with
respect to the same system with na\"{\i}ve labeling.

Tables \ref{optimization results} and \ref{optimization results +
normalization} display corresponding results for optimization.
Particularly, Table \ref{optimization results} shows the (elapsed)
time measured in milliseconds to solve the optimization problem
considered in the benchmarking process, whereas
Table~\ref{optimization results + normalization} shows the
speed-up of our system with respect to the rest of the systems.

In these tables, all numbers represent the average of ten runs.
The symbol ?? means that we did not receive a solution for the
benchmark in a reasonable time and (?) indicates a non-determined
value. The symbol {\bf N} in the PAKCS, SWI Prolog and Ciao Prolog
columns mean that we could not formulate the benchmark because of
insufficient provision for constraints.

Also the notation {\em OGS} in the SWI column indicates that we
received an error of {\em Out Of Global Stack} and, consequently, no
answer was returned. In the GNU Prolog column, the notation {\em
(SO) number} means that, in the first execution of the program no
answer was calculated because a {\em Stack Overflow} error was
raised, and that, after increasing significantly the corresponding
(cstr and trail)  environment variables, in further executions we
obtained an answer in the (average) time indicated by {\em number}.
The notation {\em RE} in the SICStus Prolog column indicates that we
also did not compute an answer because a {\em Resource Error by
Insufficient Memory} was returned. The dash (-) in the Ciao Prolog
column means that we received an incorrect answer for this
benchmark\footnote{This event seems to be caused by a bug existing
in the ${\cal FD}$ constraint package.}.

As already declared, whenever possible we maintained the same
formulation for all the benchmarks in each system. However, this
was not always possible in the magic series benchmark. In the
$\toyfd$, PAKCS and SICStus Prolog systems, this problem was coded
by using specific constraints (i.e., $count/4$, $sum/3$ and
$scalar\_product/4$ - see formulation in Example~\ref{ex: magic
series}). However, the GNU Prolog system lacks these constraints
and, therefore, we used a classical formulation that requires to
use reified constraints \cite{codognet+:clp(fd)-jlp96}. This
classical formulation is somewhat different in $\Toyfd$ since
reification applies to Boolean types (whilst in GNU Prolog, as in
general in $\Clpfd$ languages, the Boolean values $false$ and
$true$ correspond to the numerical values 0 and 1 respectively).
On the other hand, it was not possible in PAKCS as reified
constraints are not available in this system. However, since
SICStus Prolog allows reified constraints, the two formulations
were considered in this system. Then, in the SICStus column and
for the magic series benchmark row, we show between parentheses
the (elapsed solving) time associated with the reified
constraints-based formulation followed by the time associated to
the alternative formulation based on the use of specific
constraints.

In the speed-up tables, in those cases in which for a particular
system either a problem could not be expressed (e.g., for PAKCS,
SWI Prolog or Ciao Prolog), or an error was returned avoiding to
compute a first answer, or an incorrect answer was returned, we
use the symbol $\infty$  to express that our system clearly
outperforms that system since our system provides constraint
support to formulate a solution for the benchmark and compute an
answer. Also, a result {\em $\geq$ x.00} indicates that $\Toyfd$
computed an answer in 0.0 milliseconds and thus no speed-up can be
calculated; in these cases, {\em x.00} indicates that $\Toyfd$ is,
at least, {\em x} times faster than the compared system.

\subsubsection{Analysis of the Results}

The third column in Tables~\ref{naive labeling normalized results}
and~\ref{normalized results for first fail labeling}, and column 2
in Table~\ref{optimization results + normalization} show that, in
general, our implementation behaves closely to that of SICStus
Prolog in both constraint satisfaction and constraint optimization
(in fact, this is not surprising as current version of $\Toyfd$
uses SICStus Prolog ${\cal FD}$ solver) except for solving linear
equations (in these cases it is between two and four times
slower). The reason seems to be in the transformation process
previous to the invocation of the ${\cal FD}$ solver. Expressions
have to be transformed into head normal form, which means that
their arguments are also transformed into head normal form (see
Section~\ref{sect:details about implementation}). Thus, there
seems to be an overhead when expressions (such as those for linear
equations) involve a high number of arguments and sub-expressions.
This may be the same reason argued to explain the slow-down of
$\Toyfd$ in the solving of the queens benchmark via first-fail
labeling, although no appreciable slow-down was shown in the
solving via na\"{\i}ve labeling.

PAKCS is between one and three times slower than our implementation.
This is quite interesting as the PAKCS implementation is fairly
efficient and is also based on the SICStus Prolog ${\cal FD}$
library. Perhaps the reason of this slowdown with respect to
$\Toyfd$ is that PAKCS implements an alternative operational model
that also supports concurrency, and this model introduces some kind
of overhead in the solving of goals.

$\Toyfd$ also performs reasonably well compared to the other
$\Clpfd$ systems.
It clearly outperforms both Ciao Prolog's and SWI Prolog's
constraint solvers which are far, in their current versions, from
the efficiency of $\Toyfd$ in the solving of constraint
satisfaction problems (for fairness, we have to say that these
results cannot be extrapolated to the whole Ciao Prolog and SWI
Prolog systems which are quite efficient; in fact, the integer
bounds constraint solver of SWI Prolog seems to be a rather
non-optimized simple integer constraint solver that probably will
be largely improved in future versions. This same argument can be
applied to the finite domain constraint solving package currently
existing in the
Ciao Prolog system that seems to be non-mature yet).
 With respect
to GNU Prolog's constraint solver, our system behaves acceptably
well if we take into account that this solver has shown an
efficiency comparable to commercial systems. Except for the N-queens
benchmark (that seems to be particularly optimized for GNU solver)
our system is in the same order of efficiency. Moreover, it even
behaves better on scalable problems when the size of the problem
increases (e.g., in the magic series problem with na\"{\i}ve
labeling). In this sense, again with the exception of the N-queens
problem, as the instance of the problem increases, the performance
of $\Toyfd$ becomes closer to that of GNU Prolog (this result is
confirmed for both constraint satisfaction and constraint
optimization).

Further, with regard to the comparison to the other $\Cflpfd$
system, we have to say that PAKCS provides a small set of global
constraints (i.e., exactly four) as mentioned in Section~\ref{sect:
Efficiency Compared}, whereas $\Toyfd$ also gives support to
specialized constraints for particular problems such as scheduling
and placements problems. Moreover, PAKCS does not provide ${\cal
FD}$ constraints that help users to recover statistics of the
constraint solving process (e.g., number of domain prunings,
entailments detected by a constraint, backtracks due to
inconsistencies, constraint resumptions, etc) which is very useful
in practice, as $\Toyfd$ does. (For the sake of fairness, we mention
again that PAKCS supports the concurrent evaluation of constraints
which is also very convenient in practice.)

Based on the results shown in this Section, we can assure that
$\Toyfd$ is the first pure $\Cflpfd$ system that provides a wide set
of ${\cal FD}$ constraints that makes it really competitive compared
to existing $\Clpfd$ systems. These results encourage us to continue
working on our approach, and we hope to further improve the results
in a close future by means of introducing further optimizations.

\section{Related Work}
\label{sect: related work}

In addition to already cited related work, in this section we
discuss some more related work.
As already said, most of the work
to integrate constraints in the declarative programming paradigm
has been developed on $LP$
\cite{codognet+:clp(fd)-jlp96,carlsson+:open-ended-plilp97}.
However, there exist some attempts to integrate constraints in the
functional logic framework. For instance,
\cite{arenas+:cflp(r)-agp96,fraguas+:multiparadigm-system-rta99}
show how to integrate both linear constraints over real numbers
and disequality constraints in the $FLP$ language $\toy$. Also,
\cite{lux:adding-real-cons-curry-flops01} describes the addition
of linear constraints over real numbers to the $FLP$ language
Curry \cite{hanus:curry99}. Our work is guided to the ${\cal FD}$
constraint, instead of real constraints (although they are
preserved), which allows to use
non-linear constraints and adapts better to a range of ${\cal FD}$
applications.

With respect to ${\cal FD}$, the closer proposal to ours is that
described in \cite{antoy+:multi-paradigm-frocos2000} that
indicated how the integration of ${\cal FD}$ constraints in $FLP$
could be carried out. As already indicated, PAKCS is an
implementation that follows these indications.

$\Toyfd$ may also be considered from a multiparadigmatic view that
means to combine constraint programming with another paradigms in
one setting. In this context, there are some similarities with the
language
\Oz~\cite{roy+:logic-programing-context-multiparadigm-prog-tplp2003,roy+:concepts-tchniques-computer-programming-mitpress04}
as this provides salient features of $FP$ such as compositional
syntax and first-class functions, and features of $LP$ and
constraint programming including logic variables, constraints, and
programmable search mechanisms. However, \Oz is quite different to
$\Toyfd$ because of a number or reasons: (1) Oz does not provide
main features of classical functional languages such as explicit
types or curried notation; (2) functional notation is provided in
\Oz as a syntactic convenience; (3) The \Oz computation mechanism
is not based on rewriting logic as that of $\Toyfd$; (4) Oz
supports a class of lazy functions based on a demand-driven
computation but this is not an inherent feature of the language
(as in $\Toyfd$) and functions have to be made lazy explicitly
(e.g., via the concept of {\em futures}); (5) functions and
constraints are not really integrated, that is to say, they do not
have the same category as in $\Toyfd$ (i.e., constraints are
functions) and both coexist in a concurrent setting, and (6) \Oz
programs follow a far less concise program syntax than $\toyfd$.
In fact, Oz generalizes the $CLP$ and concurrent constraint
programming paradigms to provide a very flexible approach to
constraint programming very different to our proposal for
$\cflpfd$.

Also, LIFE \cite{ait-kaci+:life-system-jlp93} is an experimental
language aiming to integrate logic programming and functional
programming but, as Oz, the proposal is quite different to
$\Toyfd$ as firstly, it is considered in the framework of
object-oriented programming, and, secondly, LIFE enables the
computation over an order-sorted domain of feature trees by
allowing the equality (i.e., unification) and entailment (i.e.,
matching) constraints over order-sorted feature terms.

\begin{sloppypar}
There exist other constraint systems that share some aspects with
$\Toyfd$ although they are very different. One of those systems is
FaCiLe
\cite{barnier+:facile-a-functional-constraint-library-alp-newsletter2001},
a constraint programming library that provides
constraint solving over integer finite domains, HO functions, type
inference, strong typing, and user-defined constraints. However,
despite these similarities, FaCiLe is very different to $\Toyfd$
as it is built on top of the functional language OCaml that
provides full imperative capabilities and does not have a logical
component; also OCaml is a strict language, as opposed to lazy
ones. In fact, as \Oz, it allows the manipulation of potentially
infinite data structures by {\em explicit} delayed expressions,
but laziness is not an inherent characteristic of the resolution
mechanism as in $\toyfd$. Moreover, FaCiLe is a library and thus
it lacks programming language features.
\end{sloppypar}

Other interesting systems are OPL \cite{hentenryck+:opl-mit99} and
AMPL \cite{fourer+:ampl-sp93} that cannot be compared to our work
because they are algebraic languages which therefore are not
general programming languages. Moreover, these languages do not
benefit neither from complex terms and patterns nor from
non-determinism as $\Toyfd$ does.

Finally, we mention here another $\Cflp$ scheme proposed in the
Phd Thesis of M. Marin
\cite{marin:FLP-distributed-constraint-solving-phdthesis-2000}.
This approach introduces
$\Cflp(\cdomd,\,\mathcal{S},\,\mathcal{L})$, a family of languages
parameterized by a constraint domain $\cdomd$, a strategy
$\mathcal{S}$ which defines the cooperation of several constraint
solvers over $\cdomd$, and a constraint lazy narrowing calculus
$\mathcal{L}$ for solving constraints involving functions defined
by user given constrained rewriting rules. This approach relies on
solid work on higher-order lazy narrowing calculi and has been
implemented on top of Mathematica
\cite{marin+:mathematica-implementation-distributed-cs-system-ims99,marin+:cooperative-constraint-functional-logic-programming-ipse2000}.
Its main limitation from our viewpoint is the lack of declarative
semantics.

\begin{sloppypar}
Generally speaking, $\Toyfd$ is, from its nature, different to all
the constraints systems discussed above since $\Toyfd$ is a {\em
pure} $FLP$ language that combines characteristics of {\em pure}
$LP$ and {\em pure} $FP$ paradigms, and its operational mechanism
is the result of combining the operational methods of logic
languages (i.e., unification and resolution) and functional
languages (i.e., rewriting).
\end{sloppypar}

\section{Conclusions and Future Work}\label{Conclusions and Future Work}

In this paper we have presented $\cflpfd$, a functional logic
programming approach to ${\cal FD}$ constraint solving. $\Cflpfd$
is not only a declarative alternative to $\Clpfd$ but also extends
its capabilities with new characteristics unusual or not existing
in $\Clpfd$ such as functional and curried notation, types,
curried and higher-order functions (e.g., higher-order
constraints), constraint composition, higher-order patterns, lazy
evaluation and polymorphism, among others. As a consequence,
$\Cflpfd$ provides better tools, when compared to $\clpfd$, for a
productive declarative programming as it implicitly enables more
expressivity, due to the combination of functional, relational and
curried notation as well as type system.  Moreover, lazy
evaluation allows the use of structures hard to manage in
$\clpfd$, such as infinite lists.

A $\Cflpfd$ language is also presented by describing its syntax,
type discipline and both declarative and operational  semantics.
${\cal FD}$ constraints are integrated as functions to make them
first-class citizens and allow their use in any place where a data
can (e.g., as arguments of functions). This provides a powerful
mechanism to define higher-order constraints.

We have also reported an implementation of the $\Cflpfd$
proposal which connects a $FLP$ language to a ${\cal FD}$
constraint solver, that provides both lazy computation and ${\cal
FD}$ constraint solving. The ${\cal FD}$ solver is required to
hold termination, soundness and completeness properties. $\Toyfd$
is our implementation of the $\Cflpfd$ language previously
described, that connects the functional logic language $\Toy$ to
the efficient ${\cal FD}$ constraint solver of SICStus Prolog. The
result is that $\Toyfd$ is a lazy functional logic system with
support for ${\cal FD}$ constraint solving.

We have also explained the most important contributions by
showing the extra capabilities of $\Cflpfd$ with respect to
$\Clpfd$.  This comparison points out  the main benefits of
integrating $FLP$ and ${\cal FD}$ in a declarative language.

Moreover, we have also shown that constraint solving in $\Toyfd$
is fairly efficient as, in general, behaves closely to SICStus
Prolog, which means that the wrapping of SICStus Prolog by $\Toy$
does not increase significantly the computation time. In addition,
$\Toyfd$ clearly outperforms existing $\Clpfd$ systems such as SWI
Prolog and Ciao Prolog and also is competitive with respect to GNU
Prolog, one of the most efficient $\Clpfd$ systems. Furthermore,
$\Toyfd$ is around one and three times faster than PAKCS, its
closer $\Cflpfd$ implementation. Practical applications of
$\Toyfd$ can be found in
\cite{fernandez+:a-cflpfd-system-combi-cons-es02,fernandez+:solving-combin-cflpfd-padl03}.

Throughout the paper it should have been clear that one inherent
advantage of the $\Cflpfd$ approach is that it enables to solve
all the $\Clpfd$ applications as well as other problems closer
to the functional setting.

We claim that the integration of ${\cal FD}$ constraints into a
$FLP$ language receive benefits from both worlds, i.e., taking
functions, higher-order patterns, partial applications,
non-determinism, lazy evaluation, logical variables, and types
from $FLP$ and domain variables, constraints, and propagators from
the ${\cal FD}$ constraint programming.

In addition, we claim that the idea of interfacing a $FLP$
language and constraint solvers can be extended to other kind of
interesting constraint systems, such as non-linear constraints,
constraints over sets, or Boolean constraints, to name a few.
Observe that $\Toyfd$ can be thought of as a constraint solving
procedure integrated into a sophisticated, state-of-the-art
execution mechanism for lazy narrowing. Operationally speaking,
$\Toyfd$ compiles $\cflpfd$-programs into Prolog-programs in a
system equipped with a constraint solver. This makes both lazy
evaluation and constraint solving be inherent features of the
system.

\section*{Acknowledgment}

We thank the anonymous reviewers for their useful comments and
suggestions that helped us to improve this paper.

%
%

\bibliographystyle{acmtrans}
\bibliography{biblio}

\appendix

\section{[Proof of Theorem 1] in Page~\pageref{teorema}}
\label{appA}

The proof of theorem $1$ can be done distinguishing several cases
from the declara\-tive semantics of each primitive function symbol
given in Table $1$ and the requirements of each
constraint solver rule or failure rule in Tables $2$-$5$:\\\\
{\bf Rules of Table $2$}\\\\
We examine for example the first rule in Table $2$: $seq ~t ~s$
$\to!$ $R,$ $S$ $\myboxi$ $\sigma$ $\red_{\chi}$ $(t == s,$
$S\theta_{1}$ $\myboxi$ $\sigma\theta_{1})$ $\vee$ $(t
~\backslash= s,$ $S\theta_{2}$ $\myboxi$ $\sigma\theta_{2})$ with
$R \notin \chi$, $\theta_{1} = \{R \mapsto true\}$ and $\theta_{2}
= \{R \mapsto false\}$ (the rest of rules in Table $2$ are
analogous). We prove that $Sol_{\cdom}(seq ~t ~s$ $\to!$ $R,$ $S$
$\myboxi$ $\sigma)$ $=$ $Sol_{\cdom}(t == s,$ $S\theta_{1}$
$\myboxi$ $\sigma\theta_{1})$ $\cup$ $Sol_{\cdom}(t ~\backslash=
s,$ $S\theta_{2}$ $\myboxi$
$\sigma\theta_{2})$:\\\\
$\subseteq)$ Let $\eta$ $\in$ $Sol_{\cdom}(seq ~t ~s$ $\to!$ $R,$
$S$ $\myboxi$ $\sigma)$. By definition of $Sol_{\cdom}$ we have
$\eta$ $\in$ $Sol_{\cdom}(seq ~t ~s$ $\to!$ $R)$ and $\eta$ $\in$
$Sol_{\cdom}(S$ $\myboxi$ $\sigma)$. Since $\eta$ $\in$
$Sol_{\cdom}(seq ~t ~s$ $\to!$ $R)$ we obtain
$seq^{\cdom}~t\eta~s\eta$ $\to$ $\eta(R)$ with $\eta(R)$ total.
According to Table $1$, $\eta(R)$ must be only $true$ or $false$.
We distinguish two cases:
\begin{itemize}
\item If $\eta(R)$ $=$ $true$ then trivially $\eta$ $\in$
$Sol_{\cdom}(seq ~t ~s$ $\to!$ $true)$ or equivalently $\eta$
$\in$ $Sol_{\cdom}(t == s)$. Moreover, since $\eta(R)$ $=$ $true$
$=$ $(true)\eta$ we have $\eta$ $\in$ $Sol(\theta_1)$ and then
$\theta_1\eta$ $=$ $\eta$ (because $\eta(\theta_1(R))$ $=$
$(true)\eta$ $=$ $\eta(R)$ and $\eta(\theta_1(X))$ $=$ $\eta(X)$
for all $X$ $\neq$ $R$). Then, since $\eta$ $\in$ $Sol_{\cdom}(S$
$\myboxi$ $\sigma)$ we also have $\theta_1\eta$ $\in$
$Sol_{\cdom}(S$ $\myboxi$ $\sigma)$, or equivalently $\eta$ $\in$
$Sol_{\cdom}(S\theta_1$ $\myboxi$ $\sigma\theta_1)$. We can
conclude $\eta$ $\in$ $Sol_{\cdom}(t == s,$ $S\theta_{1}$
$\myboxi$ $\sigma\theta_{1})$.
\item If $\eta(R)$ $=$ $false$, using an analogous reasoning, we
can also conclude $\eta$ $\in$ $Sol_{\cdom}(t$ $\backslash=$ $s,$
$S\theta_{2}$ $\myboxi$ $\sigma\theta_{2})$.
\end{itemize}
Therefore, $\eta$ $\in$ $Sol_{\cdom}(t == s,$ $S\theta_{1}$
$\myboxi$ $\sigma\theta_{1})$ $\cup$ $Sol_{\cdom}(t ~\backslash=
s,$ $S\theta_{2}$ $\myboxi$ $\sigma\theta_{2})$.\\\\
$\supseteq)$ Let $\eta$ $\in$ $Sol_{\cdom}(t == s,$ $S\theta_{1}$
$\myboxi$ $\sigma\theta_{1})$ $\cup$ $Sol_{\cdom}(t ~\backslash=
s,$ $S\theta_{2}$ $\myboxi$ $\sigma\theta_{2})$. We distinguish
again two cases:
\begin{itemize}
\item If $\eta$ $\in$ $Sol_{\cdom}(t == s,$ $S\theta_{1}$
$\myboxi$ $\sigma\theta_{1})$ then, by definition of $Sol_{\cdom}$
we have $\eta$ $\in$ $Sol_{\cdom}(t == s)$ and $\eta$ $\in$
$Sol_{\cdom}(S\theta_{1}$ $\myboxi$ $\sigma\theta_{1})$ (or
equivalently, $\eta$ $\in$ $Sol_{\cdom}(S\theta_{1})$ and $\eta$
$\in$ $Sol_{\cdom}(\sigma\theta_{1})$). Since $\eta$ $\in$
$Sol_{\cdom}(\sigma\theta_{1})$ and $R$ $\notin$ $dom(\sigma)$ (by
initial hypothesis, $seq ~t ~s$ $\to!$ $R,$ $S$ $\myboxi$ $\sigma$
satisfy the requirements of Definition $2$) we deduce $\eta$ $\in$
$Sol(\theta_1)$ and then $\eta(R)$ $=$ $(true)\eta$ $=$ $true$.
But then, $\eta$ $\in$ $Sol_{\cdom}(seq ~t ~s$ $\to!$ $R)$ because
$seq^{\cdom}~t\eta~s\eta$ $\to$ $\eta(R)$ $=$ $true$ and we have
$\eta$ $\in$ $Sol_{\cdom}(t == s)$. Moreover, $\theta_1\eta$ $=$
$\eta$ (because $\eta(\theta_1(R))$ $=$ $(true)\eta$ $=$ $\eta(R)$
and $\eta(\theta_1(X))$ $=$ $\eta(X)$ for all $X$ $\neq$ $R$) and
we can also obtain $\eta$ $\in$ $Sol_{\cdom}(S$ $\myboxi$
$\sigma)$ because $\eta$ $\in$ $Sol_{\cdom}(S\theta_1$ $\myboxi$
$\sigma\theta_1)$, or equivalently, $\theta_1 \eta$ $\in$
$Sol_{\cdom}(S$ $\myboxi$ $\sigma)$. Therefore, $\eta$ $\in$
$Sol_{\cdom}(seq ~t ~s$ $\to!$ $R,$ $S$ $\myboxi$ $\sigma)$.
\item If $\eta$ $\in$ $Sol_{\cdom}(t ~\backslash= s,$
$S\theta_{2}$ $\myboxi$ $\sigma\theta_{2})$, using an analogous
reasoning, we can also conclude $\eta$ $\in$ $Sol_{\cdom}(seq ~t
~s$ $\to!$ $R,$ $S$ $\myboxi$ $\sigma)$.
\end{itemize}
The remaining conditions of the theorem for this rule trivially
hold because of the initial hypothesis $seq ~t ~s$ $\to!$ $R,$ $S$
$\myboxi$ $\sigma$ satisfies the requirements of Definition $2$, and
because of the
conditions of the rule $R$ $\notin$ $\chi$.\\\\
{\bf Rules of Table $3$}\\\\
We examine the first rule in Table $3$: $u == u,$ $S$ $\myboxi$
$\sigma$ $\red_{\chi}$ $S$ $\myboxi$ $\sigma$ with $u\in
\mathbb{Z}$. In this case, trivially $Sol_{\cdom}(u == u,$ $S$
$\myboxi$ $\sigma)$ $=$ $Sol_{\cdom}(u == u)$ $\cap$
$Sol_{\cdom}(S$ $\myboxi$ $\sigma)$ $=$ $Val(\cdom)$ $\cap$
$Sol_{\cdom}(S$ $\myboxi$ $\sigma)$ $=$ $Sol_{\cdom}(S$ $\myboxi$
$\sigma)$. The remaining conditions of the theorem trivially holds
by initial hypothesis. We examine now the second rule in Table
$3$: $X == t,$ $S$ $\myboxi$ $\sigma$ $\red_{\chi}$ $t == t,$
$S\theta$ $\myboxi$ $\sigma\theta$ with $X$ $\notin$ $\chi$ $\cup$
$var(t),$ $var(t)$ $\cap$ $\chi$ $=$ $\emptyset$ and $\theta$ $=$
$\{X \mapsto t\}$. We prove that $Sol_{\cdom}(X == t,$ $S$
$\myboxi$ $\sigma)$ $=$
$Sol_{\cdom}(t == t,$ $S\theta$ $\myboxi$ $\sigma\theta)$:\\\\
$\subseteq)$ Let $\eta$ $\in$ $Sol_{\cdom}(X == t,$ $S$ $\myboxi$
$\sigma)$. By definition of $Sol_{\cdom}$ we have $\eta$ $\in$
$Sol_{\cdom}(X == t)$ and $\eta$ $\in$ $Sol_{\cdom}(S$ $\myboxi$
$\sigma)$. Since $\eta$ $\in$ $Sol_{\cdom}(X == t)$ we obtain
$seq^{\cdom}~\eta(X)~t\eta$ $\to$ $true$. According to Table $1$
we obtain $\eta(X)$ $=$ $t\eta$ with $t\eta$ total and then $\eta$
$\in$ $Sol(\theta)$. In this situation, trivially $\eta$ $\in$
$Sol_{\cdom}(t == t)$. Moreover, since $\eta$ $\in$ $Sol(\theta)$,
we deduce $\theta\eta$ $=$ $\eta$ (because $\eta(\theta(X))$ $=$
$t\eta$ $=$ $\eta(X)$ and $\eta(\theta(Y))$ $=$ $\eta(Y)$ for all
$Y$ $\neq$ $X$). Then, since $\eta$ $\in$ $Sol_{\cdom}(S$
$\myboxi$ $\sigma)$, we also have $\theta\eta$ $\in$
$Sol_{\cdom}(S$ $\myboxi$ $\sigma)$, or equivalently $\eta$ $\in$
$Sol_{\cdom}(S\theta$ $\myboxi$ $\sigma\theta)$. Therefore, we can
conclude $\eta$ $\in$
$Sol_{\cdom}(t == t,$ $S\theta$ $\myboxi$ $\sigma\theta)$.\\\\
$\supseteq)$ Let $\eta$ $\in$ $Sol_{\cdom}(t == t,$ $S\theta$
$\myboxi$ $\sigma\theta)$. By definition of $Sol_{\cdom}$ we have
$\eta$ $\in$ $Sol_{\cdom}(t == t)$ and $\eta$ $\in$
$Sol_{\cdom}(S\theta$ $\myboxi$ $\sigma\theta)$ (or equivalently,
$\eta$ $\in$ $Sol_{\cdom}(S\theta)$ and $\eta$ $\in$
$Sol_{\cdom}(\sigma\theta)$). Since $\eta$ $\in$
$Sol_{\cdom}(\sigma\theta)$ and $X$ $\notin$ $dom(\sigma)$ (by
initial hypothesis, $X == t,$ $S$ $\myboxi$ $\sigma$ satisfies the
requirements of Definition $2$) we deduce $\eta$ $\in$
$Sol(\theta)$ and then $\eta(X)$ $=$ $t\eta$. But then, $\eta$
$\in$ $Sol_{\cdom}(X == t)$ because $\eta$ $\in$ $Sol_{\cdom}(t ==
t)$ and $seq^{\cdom}~\eta(X)~t\eta$ $\to$ $true$ with $\eta(X)$
$=$ $t\eta$ total. Moreover, $\theta\eta$ $=$ $\eta$ (because
$\eta(\theta(X))$ $=$ $t\eta$ $=$ $\eta(X)$ and $\eta(\theta(Y))$
$=$ $\eta(Y)$ for all $Y$ $\neq$ $X$) and we can obtain $\eta$
$\in$ $Sol_{\cdom}(S$ $\myboxi$ $\sigma)$ because $\eta$ $\in$
$Sol_{\cdom}(S\theta$ $\myboxi$ $\sigma\theta)$, or equivalently,
$\theta\eta$ $\in$ $Sol_{\cdom}(S$ $\myboxi$ $\sigma)$. Therefore
$\eta$ $\in$
$Sol_{\cdom}(X == t,$ $S$ $\myboxi$ $\sigma)$.\\\\
The remaining conditions of the theorem for this rule trivially
hold because of the initial hypothesis $X == t,$ $S$ $\myboxi$
$\sigma$ satisfies the requirements of Definition $2$, and because of the
conditions of the rule $X$ $\notin$ $\chi$ $\cup$ $var(t)$ and
$var(t)$ $\cap$ $\chi$ $=$ $\emptyset$. Finally, we examine the
main rule in Table $3$ for strict disequality (the rest of rules
in Table $3$ are ana\-logous or more simples): $X$ $\backslash=$
$h~\tpp{t}{n},$ $S$ $\myboxi$ $\sigma$ $\red_{\chi}$
$(\bigvee_{i}(S\theta_{i}$ $\myboxi$ $\sigma\theta_{i}))$ $\vee$
$(\bigvee_{k=1}^{n}(U_{k} ~\backslash= t_{k}\theta,$ $S\theta$
$\myboxi$ $\sigma\theta))$ with $X$ $\notin$ $\chi,$
$var(h~\tpp{t}{n})$ $\cap$ $\chi$ $\neq$ $\emptyset$, $\theta_{i}$
$=$ $\{X \mapsto$ $h_{i}$ $\tpp{Y}{m_i}\}$ with $h_{i}$ $\neq$
$h$, and $\theta$ $=$ $\{X \mapsto h$ $\tpp{U}{n}\}$ with
$\overline{Y}_{m_{i}},$ $\overline{U}_{n}$ new fresh variables. We
prove that $Sol_{\cdom}(X$ $\backslash=$ $h~\tpp{t}{n},$ $S$
$\myboxi$ $\sigma)$ $=$ $($ $\bigcup_i$ $Sol_{\cdom}(\exists
\tpp{Y}{m_i}.$ $(S\theta_{i}$ $\myboxi$ $\sigma\theta_{i}))$ $)$
$\cup$ $($ $\bigcup_{k=1}^n$ $Sol_{\cdom}(\exists \tpp{U}{n}.$
$(U_{k} ~\backslash= t_{k}\theta,$
$S\theta$ $\myboxi$ $\sigma\theta))$ $)$:\\\\
$\subseteq)$ Let $\eta$ $\in$ $Sol_{\cdom}(X$ $\backslash=$
$h~\tpp{t}{n},$ $S$ $\myboxi$ $\sigma)$. By definition of
$Sol_{\cdom}$ we have $\eta$ $\in$ $Sol_{\cdom}(X$ $\backslash=$
$h~\tpp{t}{n})$ and $\eta$ $\in$ $Sol_{\cdom}(S$ $\myboxi$
$\sigma)$. Since $\eta$ $\in$ $Sol_{\cdom}(X$ $\backslash=$
$h~\tpp{t}{n})$ we obtain $seq^{\cdom}$ $\eta(X)$
$(h~\tpp{t}{n})\eta$ $\to$ $false$. According to Table $1$,
$\eta(X)$ and $(h~\tpp{t}{n})\eta$ $=$ $h~\overline{t_n\eta}$ have
no common upper bound w.r.t. the information ordering
$\sqsubseteq$, and we can distinguish two cases:
\begin{itemize}
\item $\eta(X)$ $=$ $h_i~\tpp{s}{m_i}$ with $h_i$ $\neq$ $h$.
Since $\tpp{Y}{m_i}$ are new variables, we can define $\eta'$
$=_{\backslash \tpp{Y}{m_i}}$ $\eta$ such that $\eta'(Y_{k})$ $=$
$s_{k}$ for all $1$ $\leq$ $k$ $\leq$ $m_i$ and $\eta'(Z)$ $=$
$\eta(Z)$ for all $Z$ $\notin$ $\tpp{Y}{m_i}$. Clearly, $\eta'(X)$
$=$ $\eta(X)$ $=$ $h_i~\tpp{s}{m_i}$ $=$
$h_i~\overline{\eta'(Y_{m_i})}$ $=$ $(h_i~\tpp{Y}{m_i})\eta'$ and
then $\eta'$ $\in$ $Sol(\theta_i)$. Moreover, $\theta_i\eta'$
$=_{\backslash \tpp{Y}{m_i}}$ $\eta$ because $\eta'(\theta_i(X))$
$=$ $(h_i~\tpp{Y}{m_i})\eta'$ $=$ $h_i~\overline{\eta'(Y_{m_i})}$
$=$ $h_i~\tpp{s}{m_i}$ $=$ $\eta(X)$ and $\eta'(\theta_i(Z))$ $=$
$\eta'(Z)$ $=$ $\eta(Z)$ for all $Z$ $\notin$ $\{X\}$ $\cup$
$\tpp{Y}{m_i}$. Since $\eta$ $\in$ $Sol_{\cdom}(S$ $\myboxi$
$\sigma)$ and $\tpp{Y}{m_i}$ are new variables in $S$ $\myboxi$
$\sigma$, we also have $\theta_i\eta'$ $\in$ $Sol_{\cdom}(S$
$\myboxi$ $\sigma)$, or equivalently, $\eta'$ $\in$
$Sol_{\cdom}(S\theta_i$ $\myboxi$ $\sigma\theta_i)$. Finally,
since there exists $\eta'$ $=_{\backslash \tpp{Y}{m_i}}$ $\eta$
with $\tpp{Y}{m_i}$ new variables such that $\eta'$ $\in$
$Sol_{\cdom}(S\theta_i$ $\myboxi$ $\sigma\theta_i)$ we can deduce
$\eta$ $\in$ $Sol_{\cdom}(\exists \tpp{Y}{m_i}.$ $(S\theta_i$
$\myboxi$ $\sigma\theta_i))$ for any $i$ such that $h_i$ $\neq$
$h$.
\item $\eta(X)$ $=$ $h~\tpp{s}{n}$ with a pattern $s_k$ $(1$
$\leq$ $k$ $\leq$ $n)$ such that $s_k$ and $t_k\eta$ have no
common upper bound w.r.t. the information ordering $\sqsubseteq$
(i.e., $seq^{\cdom}$ $s_k$ $t_k\eta$ $\to$ $false$). Since
$\tpp{U}{n}$ are new variables, we can define $\eta'$
$=_{\backslash \tpp{U}{n}}$ $\eta$ such that $\eta'(U_k)$ $=$
$s_k$ for all $1$ $\leq$ $k$ $\leq$ $n$ and $\eta'(Y)$ $=$
$\eta(Y)$ for all $Y$ $\notin$ $\tpp{U}{n}$. Clearly, $\eta'(X)$
$=$ $\eta(X)$ $=$ $h~\tpp{s}{n}$ $=$ $h~\overline{\eta'(U_n)}$ $=$
$(h~\tpp{U}{n})\eta'$ and then $\eta'$ $\in$ $Sol(\theta)$.
Moreover, $\theta\eta'$ $=_{\backslash \tpp{U}{n}}$ $\eta$ because
$\eta'(\theta(X))$ $=$ $(h~\tpp{U}{n})\eta'$ $=$
$h~\overline{\eta'(U_n)}$ $=$ $h~\tpp{s}{n}$ $=$ $\eta(X)$ and
$\eta'(\theta(Y))$ $=$ $\eta'(Y)$ $=$ $\eta(Y)$ for all $Y$
$\notin$ $\{X\}$ $\cup$ $\tpp{U}{n}$. Therefore, there exists $1$
$\leq$ $k$ $\leq$ $n$ such that $seq^{\cdom}$ $\eta'(U_k)$
$t_k\theta\eta'$ $\to$ $false$ because $\eta'(U_k)$ $=$ $s_k$ and
$t_k\theta\eta'$ $=$ $t_k\eta$ (since $\tpp{U}{n}$ are new
variables, $var(t_k)$ $\cap$ $\tpp{U}{n}$ $=$ $\emptyset$) and we
can deduce $\eta'$ $\in$ $Sol_{\cdom}(U_k$ $\backslash=$
$t_k\theta)$. On the other hand, $\eta$ $\in$ $Sol_{\cdom}(S$
$\myboxi$ $\sigma)$, or equivalently $\theta\eta'$ $\in$
$Sol_{\cdom}(S$ $\myboxi$ $\sigma)$, because $\tpp{U}{n}$ are
again new variables in $S$ $\myboxi$ $\sigma$. We can also
conclude $\eta'$ $\in$ $Sol_{\cdom}(S\theta$ $\myboxi$
$\sigma\theta)$. Finally, since there exists $\eta'$
$=_{\backslash \tpp{U}{n}}$ $\eta$ with $\tpp{U}{n}$ new variables
such that $\eta'$ $\in$ $Sol_{\cdom}(U_k$ $\backslash=$
$t_k\theta,$ $S\theta$ $\myboxi$ $\sigma\theta)$, we obtain $\eta$
$\in$ $Sol_{\cdom}(\exists \tpp{U}{n}.$ $(U_k$ $\backslash=$
$t_k\theta,$ $S\theta$ $\myboxi$ $\sigma\theta))$ $(1$ $\leq$ $k$
$\leq$ $n)$.
\end{itemize}
$\supseteq)$ Let $\eta$ $\in$ $($ $\bigcup_i$ $Sol_{\cdom}(\exists
\tpp{Y}{m_i}.$ $(S\theta_{i}$ $\myboxi$ $\sigma\theta_{i}))$ $)$
$\cup$ $($ $\bigcup_{k=1}^n$ $Sol_{\cdom}(\exists \tpp{U}{n}.$
$(U_{k} ~\backslash= t_{k}\theta,$ $S\theta$ $\myboxi$
$\sigma\theta))$ $)$. We distinguish again two cases:
\begin{itemize}
\item $\eta$ $\in$ $Sol_{\cdom}(\exists \tpp{Y}{m_i}.$
$(S\theta_{i}$ $\myboxi$ $\sigma\theta_{i}))$ for any $i$ such
that $h_i$ $\neq$ $h$. By definition of $Sol_{\cdom}$, there
exists $\eta'$ $=_{\backslash \tpp{Y}{m_i}}$ $\eta$ such that
$\eta'$ $\in$ $Sol_{\cdom}(S\theta_{i}$ $\myboxi$
$\sigma\theta_{i})$ (or equiva\-lently, $\eta'$ $\in$
$Sol_{\cdom}(S\theta_{i})$ and $\eta'$ $\in$
$Sol_{\cdom}(\sigma\theta_{i})$). Since $\eta'$ $\in$
$Sol_{\cdom}(\sigma\theta_{i})$ and $X$ $\notin$ $dom(\sigma)$ (by
initial hypothesis, $X$ $\backslash=$ $h~\tpp{t}{n},$ $S$
$\myboxi$ $\sigma$ satisfies the requirements of Definition $2$), we
deduce $\eta'$ $\in$ $Sol(\theta_i)$ and then $\eta'(X)$ $=$
$(h_i~\tpp{Y}{m_i})\eta'$ $=$ $h_i~\overline{\eta'(Y_{m_i})}$.
Moreover, since $\eta'$ $=_{\backslash \tpp{Y}{m_i}}$ $\eta$, we
also deduce $\theta_i\eta'$ $=_{\backslash \tpp{Y}{m_i}}$ $\eta$
because $\eta'(\theta_i(X))$ $=$ $(h_i~\tpp{Y}{m_i})\eta'$ $=$
$\eta'(X)$ $=$ $\eta(X)$ and $\eta'(\theta_i(Z))$ $=$ $\eta'(Z)$
$=$ $\eta(Z)$ for all $Z$ $\notin$ $\{X\}$ $\cup$ $\tpp{Y}{m_i}$.
In this situation, $seq^{\cdom}$ $\eta'(X)$ $(h~\tpp{t}{n})\eta'$
$\to$ $false$, because $\eta'(X)$ $=$ $(h_i~\tpp{Y}{m_i})\eta'$
$=$ $h_i~\overline{\eta'(Y_{m_i})}$ and $(h~\tpp{t}{n})\eta'$ $=$
$h~\overline{t_{n}\eta'}$ with $h_i$ $\neq$ $h$ have no common
upper bound w.r.t. the information ordering $\sqsubseteq$.
Therefore, $\eta'$ $\in$ $Sol_{\cdom}(X$ $\backslash=$
$h~\tpp{t}{n})$, and we also have $\eta$ $\in$ $Sol_{\cdom}(X$
$\backslash=$ $h~\tpp{t}{n})$ because $\tpp{Y}{m_i}$ are new
variables in $X$ $\backslash=$ $h~\tpp{t}{n}$ and $\eta'$
$=_{\backslash \tpp{Y}{m_i}}$ $\eta$. On the other hand, since
$\eta'$ $\in$ $Sol_{\cdom}(S\theta_{i}$ $\myboxi$
$\sigma\theta_{i})$, or equivalently $\theta_i\eta'$ $\in$
$Sol_{\cdom}(S$ $\myboxi$ $\sigma)$, and $\tpp{Y}{m_i}$ are new
variables in $S$ $\myboxi$ $\sigma$, we obtain $\eta$ $\in$
$Sol_{\cdom}(S$ $\myboxi$ $\sigma)$ because $\theta_i\eta'$
$=_{\backslash \tpp{Y}{m_i}}$ $\eta$. Therefore, $\eta$ $\in$
$Sol_{\cdom}(X$ $\backslash=$ $h~\tpp{t}{n},$ $S$ $\myboxi$
$\sigma)$.
\item $\eta$ $\in$ $Sol_{\cdom}(\exists \tpp{U}{n}.$ $(U_{k}
~\backslash= t_{k}\theta,$ $S\theta$ $\myboxi$ $\sigma\theta))$
$(1$ $\leq$ $k$ $\leq$ $n)$. By definition of $Sol_{\cdom}$, there
exists $\eta'$ $=_{\backslash \tpp{U}{n}}$ $\eta$ such that
$\eta'$ $\in$ $Sol_{\cdom}(U_{k} ~\backslash= t_{k}\theta,$
$S\theta$ $\myboxi$ $\sigma\theta)$ $(1$ $\leq$ $k$ $\leq$ $n)$.
By definition of $Sol_{\cdom}$ again we have $\eta'$ $\in$
$Sol_{\cdom}(U_{k} ~\backslash= t_{k}\theta)$ and $\eta'$ $\in$
$Sol_{\cdom}(S\theta$ $\myboxi$ $\sigma\theta)$ (or equivalently,
$\eta'$ $\in$ $Sol_{\cdom}(S\theta)$ and $\eta'$ $\in$
$Sol_{\cdom}(\sigma\theta)$). Since $\eta'$ $\in$
$Sol_{\cdom}(\sigma\theta)$ and $X$ $\notin$ $dom(\sigma)$ (by
initial hypothesis, $X$ $\backslash=$ $h~\tpp{t}{n},$ $S$
$\myboxi$ $\sigma$ satisfies the requirements of Definition $2$) we
deduce $\eta'$ $\in$ $Sol(\theta)$ and then $\eta'(X)$ $=$
$(h~\tpp{U}{n})\eta'$ $=$ $h~\overline{\eta'(U_n)}$. Moreover,
since $\eta'$ $=_{\backslash \tpp{U}{n}}$ $\eta$, we also deduce
$\theta\eta'$ $=_{\backslash \tpp{U}{n}}$ $\eta$ because
$\eta'(\theta(X))$ $=$ $(h~\tpp{U}{n})\eta'$ $=$ $\eta'(X)$ $=$
$\eta(X)$ and $\eta'(\theta(Z))$ $=$ $\eta'(Z)$ $=$ $\eta(Z)$ for
all $Z$ $\notin$ $\{X\}$ $\cup$ $\tpp{U}{n}$. Since $\eta'$ $\in$
$Sol_{\cdom}(U_{k} ~\backslash= t_{k}\theta)$, and according to
Table $1$, we have $seq^{\cdom}$ $\eta'(U_k)$ $t_k\theta\eta'$
$\to$ $false$ where $\eta'(U_k)$ and $t_k\theta\eta'$ $=$
$t_k\eta$ (because $var(t_k)$ $\cap$ $\tpp{U}{n}$ $=$ $\emptyset$)
have no common upper bound w.r.t. the information ordering
$\sqsubseteq$. In this situation, we also have $seq^{\cdom}$
$\eta(X)$ $(h~\tpp{t}{n})\eta$ $\to$ $false$ because $\eta(X)$ $=$
$(h~\tpp{U}{n})\eta'$ $=$ $h~\overline{\eta'(U_n)}$,
$(h~\tpp{t}{n})\eta$ $=$ $h~\overline{t_n\eta}$, and clearly
$\eta(X)$ and $(h~\tpp{t}{n})\eta$ have no common upper bound
w.r.t. the information ordering $\sqsubseteq$ (there exists $1$
$\leq$ $k$ $\leq$ $n$ such that $\eta'(U_k)$ and $t_k\eta$ have no
common upper bound w.r.t. the information ordering $\sqsubseteq$).
Therefore, $\eta$ $\in$ $Sol_{\cdom}(X$ $\backslash =$
$h~\tpp{t}{n})$. On the other hand, since $\eta'$ $\in$
$Sol_{\cdom}(S\theta$ $\myboxi$ $\sigma\theta)$, or equivalently
$\theta\eta'$ $\in$ $Sol_{\cdom}(S$ $\myboxi$ $\sigma)$, and
$\tpp{U}{n}$ are new variables in $S$ $\myboxi$ $\sigma$, we
obtain $\eta$ $\in$ $Sol_{\cdom}(S$ $\myboxi$ $\sigma)$ because
$\theta\eta'$ $=_{\backslash \tpp{U}{n}}$ $\eta$. Therefore,
$\eta$ $\in$ $Sol_{\cdom}(X$ $\backslash=$ $h~\tpp{t}{n},$ $S$
$\myboxi$ $\sigma)$.
\end{itemize}
The remaining conditions of the theorem for this rule trivially
hold because of the initial hypothesis $X$ $\backslash =$
$h~\tpp{t}{n},$ $S$ $\myboxi$ $\sigma$ satisfies the requirements of
Definition $2$, and because of the conditions of the rule $X$ $\notin$
$\chi,$ $var(h~\tpp{t}{n})$ $\cap$ $\chi$ $\neq$ $\emptyset$, and
$\overline{Y}_{m_{i}},$
$\overline{U}_{n}$ are new fresh variables.\\\\
{\bf Rules of Table $4$}\\\\
We examine the first rule in Table $4$: $u \leq u',$ $S$
$\myboxi$ $\sigma$ $\red_{\chi}$ $S$ $\myboxi$ $\sigma$ with
$u,u'$ $\in$ $\mathbb{Z}$ and $u$ $\leq^{\mathbb{Z}}$ $u'$. In
this case, trivially $Sol_{\cdom}(u$ $\leq$ $u',$ $S$ $\myboxi$
$\sigma)$ $=$ $Sol_{\cdom}(u$ $\leq$ $u')$ $\cap$ $Sol_{\cdom}(S$
$\myboxi$ $\sigma)$ $=$ $Val(\cdom)$ $\cap$ $Sol_{\cdom}(S$
$\myboxi$ $\sigma)$ $=$ $Sol_{\cdom}(S$ $\myboxi$ $\sigma)$. The
remaining conditions of the theorem trivially hold by the initial
hypothesis. We examine now the main rule in Table $4$ (the rest
of rules are analogous or more simples): $a$ $\otimes$ $b$ $=$
$X,$ $S$ $\myboxi$ $\sigma$ $\red_{\chi}$ $S\theta$ $\myboxi$
$\sigma\theta$ with $X$ $\notin$ $\chi,$ $a, b$ $\in$ $\mathbb{Z}$
and $\theta$ $=$ $\{X$ $\mapsto$ $a$ $\otimes^{\mathbb{Z}}$ $b\}$.
We prove that $Sol_{\cdom}(a$ $\otimes$ $b$ $=$ $X,$ $S$ $\myboxi$
$\sigma)$ $=$
$Sol_{\cdom}(S\theta$ $\myboxi$ $\sigma\theta)$:\\\\
$\subseteq)$ Let $\eta$ $\in$ $Sol_{\cdom}(a$ $\otimes$ $b$ $=$
$X,$ $S$ $\myboxi$ $\sigma)$. By definition of $Sol_{\cdom}$ we
have $\eta$ $\in$ $Sol_{\cdom}(a$ $\otimes$ $b$ $=$ $X)$ and
$\eta$ $\in$ $Sol_{\cdom}(S$ $\myboxi$ $\sigma)$. Since $\eta$
$\in$ $Sol_{\cdom}(a$ $\otimes$ $b$ $=$ $X)$ we obtain
$\otimes^{\cdom}$ $a$ $b$ $\to$ $a$ $\otimes^{\mathbb{Z}}$ $b$,
$seq^{\cdom}$ $(a$ $\otimes^{\mathbb{Z}}$ $b)$ $\eta(X)$ $\to$
$true$ where $a,$ $b,$ $a$ $\otimes^{\mathbb{Z}}$ $b$ $\in$
$\mathbb{Z}$. According to Table $1$, we obtain $\eta(X)$ $=$ $a$
$\otimes^{\mathbb{Z}}$ $b$ $=$ $(a$ $\otimes^{\mathbb{Z}}$
$b)\eta$ and then $\eta$ $\in$ $Sol(\theta)$. Moreover, we deduce
$\theta\eta$ $=$ $\eta$ because $\eta(\theta(X))$ $=$ $(a$
$\otimes^{\mathbb{Z}}$ $b)\eta$ $=$ $a$ $\otimes^{\mathbb{Z}}$ $b$
$=$ $\eta(X)$ and $\eta(\theta(Y))$ $=$ $\eta(Y)$ for all $Y$
$\neq$ $X$. Since $\eta$ $\in$ $Sol_{\cdom}(S$ $\myboxi$ $\sigma)$
we also have $\theta\eta$ $\in$ $Sol_{\cdom}(S$ $\myboxi$
$\sigma)$, or equivalently, $\eta$ $\in$
$Sol_{\cdom}(S\theta$ $\myboxi$ $\sigma\theta)$.\\\\
$\supseteq)$ Let $\eta$ $\in$ $Sol_{\cdom}(S\theta$ $\myboxi$
$\sigma\theta)$. By definition of $Sol_{\cdom}$ we have $\eta$
$\in$ $Sol_{\cdom}(S\theta)$ and $\eta$ $\in$
$Sol_{\cdom}(\sigma\theta)$. Since by initial hypothesis $a$
$\otimes$ $b$ $=$ $X,$ $S$ $\myboxi$ $\sigma$ satisfies the
requirements of Definition $2$, we have $X$ $\notin$ $dom(\sigma)$
and then $\eta$ $\in$ $Sol(\theta)$ (i.e., $\eta(X)$ $=$ $(a$
$\otimes^{\mathbb{Z}}$ $b)\eta$ $=$ $(a$ $\otimes^{\mathbb{Z}}$
$b)$ $\in$ $\mathbb{Z}$, where $a,$ $b$ $\in$ $\mathbb{Z}$).
But then $\otimes^{\cdom}$ $a$ $b$ $\to$ $a$
$\otimes^{\mathbb{Z}}$ $b$, $seq^{\cdom}$ $(a$
$\otimes^{\mathbb{Z}}$ $b)$ $\eta(X)$ $\to$ $true$, and therefore
$\eta$ $\in$ $Sol_{\cdom}(a$ $\otimes$ $b$ $=$ $X)$. Moreover,
$\theta\eta$ $=$ $\eta$ because $\eta(\theta(X))$ $=$ $(a$
$\otimes^{\mathbb{Z}}$ $b)\eta$ $=$ $a$ $\otimes^{\mathbb{Z}}$ $b$
$=$ $\eta(X)$ and $\eta(\theta(Y))$ $=$ $\eta(Y)$ for all $Y$
$\neq$ $X$. Since $\eta$ $\in$ $Sol_{\cdom}(S\theta$ $\myboxi$
$\sigma\theta)$, or equivalently $\theta\eta$ $\in$
$Sol_{\cdom}(S$ $\myboxi$ $\sigma)$, we obtain $\eta$ $\in$
$Sol_{\cdom}(S$ $\myboxi$ $\sigma)$. Therefore, $\eta$ $\in$
$Sol_{\cdom}(a$
$\otimes$ $b$ $=$ $X,$ $S$ $\myboxi$ $\sigma)$.\\\\
The remaining conditions of the theorem for this rule trivially
hold because of the initial hypothesis $a$ $\otimes$ $b$ $=$ $X,$ $S$
$\myboxi$ $\sigma$ satisfies the requirements of Definition $2$, and
because of
the conditions of the rule $X$ $\notin$ $\chi$.\\\\
{\bf Rules of Table $5$}\\\\
We examine the first rule in Table $5$: $u$ $\in$ $[u_{1},$
$\ldots,$ $u_{n}],$ $S$ $\myboxi$ $\sigma$ $\red_{\chi}$ $S$
$\myboxi$ $\sigma$ with $u,$ $u_{i}$ $\in$ $\mathbb{Z}$ $\cup$
$\Var$ and $\exists i$ $\in$ $\{1,$ $\ldots,$ $n\}.$ $u_{i}$
$\equiv$ $u$. In this situation, and according to Table $1$, we
have $Sol_{\cdom}(u$ $\in$ $[u_{1},$ $\ldots,$ $u_{n}])$ $=$
$Val(\cdom)$: $\eta$ $\in$ $Sol_{\cdom}(u$ $\in$ $[u_{1},$
$\ldots,$ $u_{n}])$ implies that $domain^{\cdom}$ $u\eta$
$[u_{1}\eta,$ $\ldots,$ $u_{n}\eta]$ $\to$ $true$ where $\forall
i$ $\in$ $\{1,$ $\ldots,$ $n\!-\!1 \}.$ $u_{i}\eta$
$\leq^{\mathbb{Z}}$ $u_{i+1}\eta$ and $\exists i$ $\in$ $\{1,$
$\ldots,$ $n \}.$ $u\eta$ $=^{\mathbb{Z}}$ $u_{i}\eta$. It holds
for all $\eta$ $\in$ $Val(\cdom)$ because of the initial hypothesis
$u$ $\in$ $[u_{1},$ $\ldots,$ $u_{n}],$ $S$ $\myboxi$ $\sigma$
satisfies the requirements of Definition $2$ (i.e., $[u_{1},$
$\ldots,$ $u_{n}]$ represents an increasing integer list), and because of
the conditions of this rule (i.e., $\exists i$ $\in$ $\{1,$
$\ldots,$ $n\}.$ $u_{i}$ $\equiv$ $u$). Then, trivially
$Sol_{\cdom}(u$ $\in$ $[u_{1},$ $\ldots,$ $u_{n}],$ $S$ $\myboxi$
$\sigma)$ $=$ $Sol_{\cdom}(u$ $\in$ $[u_{1},$ $\ldots,$ $u_{n}])$
$\cap$ $Sol_{\cdom}(S$ $\myboxi$ $\sigma)$ $=$ $Val(\cdom)$ $\cap$
$Sol_{\cdom}(S$ $\myboxi$ $\sigma)$ $=$ $Sol_{\cdom}(S$ $\myboxi$
$\sigma)$. The remaining conditions of the theorem for this rule
trivially hold by the initial hypothesis. The second rule in
Table $5$ is completely analogous: $Sol_{\cdom}(u$ $\notin$
$[u_{1}, \ldots,u_{n}])$ $=$ $Val(\cdom)$ because $u,$ $u_{i}$
$\in$ $\mathbb{Z}$, $\forall i$ $\in$ $\{1,$ $\ldots,$ $n\}.$
$u_{i}$ $\neq^{\mathbb{Z}}$ $u$, and according to Table $1$,
$domain^{\cdom}$ $u\eta$ $[u_1\eta,$ $\ldots,$ $u_n\eta]$ $\to$
$false$ holds for all $\eta$ $\in$ $Val(\cdom)$.\\\\
Finally, we examine the main rule for $labeling$ in Table $5$:
$labeling$ $[\ldots]$ $[X],$ $X$ $\in$ $[u_{1},$ $\ldots,$
$u_{n}],$ $S$ $\myboxi$ $\sigma$ $\red_{\chi}$ $\bigvee_{i=1}^{n}$
$(S\theta_{i}$ $\myboxi$ $\sigma\theta_{i})$ with $X$ $\notin$
$\chi$, and $\forall i$ $\in$ $\{1,\ldots, n\}$, $u_{i}$ $\in$
$\mathbb{Z}$, $\theta_{i}$ $=$ $\{X \mapsto u_{i}\}$. We prove
that $Sol_{\cdom}(labeling$ $[\ldots]$ $[X],$ $X$ $\in$ $[u_{1},$
$\ldots,$ $u_{n}],$ $S$ $\myboxi$ $\sigma)$ $=$
$\bigcup_{i=1}^{n}$
$Sol_{\cdom}(S\theta_{i}$ $\myboxi$ $\sigma\theta_{i})$:\\\\
$\subseteq)$ Let $\eta$ $\in$ $Sol_{\cdom}(labeling$ $[\ldots]$
$[X],$ $X$ $\in$ $[u_{1},$ $\ldots,$ $u_{n}],$ $S$ $\myboxi$
$\sigma)$. By definition of $Sol_{\cdom}$ we have $\eta$ $\in$
$Sol_{\cdom}(labeling$ $[\ldots]$ $[X],$ $X$ $\in$ $[u_{1},$
$\ldots,$ $u_{n}])$ and $\eta$ $\in$ $Sol_{\cdom}(S$ $\myboxi$
$\sigma)$. Then, $indomain^{\cdom}$ $\eta(X)$ $\to$ $\top$,
$domain^{\cdom}$ $\eta(X)$ $[u_1,$ $\ldots,$ $u_n]$ $\to$ $true$
because $u_i$ $\in$ $\mathbb{Z}$ for all $1$ $\leq$ $i$ $\leq$
$n$. According to Table $1$, we deduce $\eta(X)$ $\in$
$\mathbb{Z}$, $\forall i$ $\in$ $\{1,$ $\ldots,$ $n\!-\!1 \} .$
$u_{i}$ $\leq^{\mathbb{Z}}$ $u_{i+1}$ and $\exists i$ $\in$ $\{1,$
$\ldots,$ $n \}.$ $\eta(X)$ $=^{\mathbb{Z}}$ $u_{i}$. Therefore,
$\eta(X)$ $=$ $u_i$ $=$ $u_i\eta$ and then $\eta$ $\in$
$Sol(\theta_i)$ $(1$ $\leq$ $i$ $\leq$ $n)$. Moreover, we have
$\theta_i\eta$ $=$ $\eta$ because $\eta(\theta_i(X))$ $=$
$u_i\eta$ $=$ $u_i$ $=$ $\eta(X)$ and $\eta(\theta_i(Y))$ $=$
$\eta(Y)$ for all $Y$ $\neq$ $X$. Finally, since $\eta$ $\in$
$Sol_{\cdom}(S$ $\myboxi$ $\sigma)$ we can conclude $\theta_i\eta$
$\in$ $Sol_{\cdom}(S$ $\myboxi$ $\sigma)$ or equivalently $\eta$
$\in$ $Sol_{\cdom}(S\theta_i$
$\myboxi$ $\sigma\theta_i)$ $(1$ $\leq$ $i$ $\leq n)$.\\\\
$\supseteq)$ Let $\eta$ $\in$ $Sol_{\cdom}(S\theta_i$ $\myboxi$
$\sigma\theta_i)$ $(1$ $\leq$ $i$ $\leq n)$. By definition of
$Sol_{\cdom}$ we have $\eta$ $\in$ $Sol_{\cdom}(S\theta_i)$ and
$\eta$ $\in$ $Sol_{\cdom}(\sigma\theta_i)$. By the initial
hypothesis $labeling$ $[\ldots]$ $[X],$ $X$ $\in$ $[u_{1},$
$\ldots,$ $u_{n}],$ $S$ $\myboxi$ $\sigma$ satisfies the
requirements of Definition $2$, we have $X$ $\notin$ $dom(\sigma)$
and then $\eta$ $\in$ $Sol(\theta_i)$ (i.e., $\eta(X)$ $=$
$u_i\eta$ $=$ $u_i$ due to $u_i$ $\in$ $\mathbb{Z}$). Moreover, we
have $\theta_i\eta$ $=$ $\eta$ because $\eta(\theta_i(X))$ $=$
$u_i\eta$ $=$ $u_i$ $=$ $\eta(X)$ and $\eta(\theta_i(Y))$ $=$
$\eta(Y)$ for all $Y$ $\neq$ $X$. Then, since $\eta$ $\in$
$Sol_{\cdom}(S\theta_i$ $\myboxi$ $\sigma\theta_i)$, or
equivalently, $\theta_i\eta$ $\in$ $Sol_{\cdom}(S$ $\myboxi$
$\sigma)$, we deduce $\eta$ $\in$ $Sol_{\cdom}(S$ $\myboxi$
$\sigma)$. Finally, we prove that $\eta$ $\in$
$Sol_{\cdom}(labeling$ $[\ldots]$ $[X],$ $X$ $\in$ $[u_{1},$
$\ldots,$ $u_{n}])$. Since $\eta(X),$ $u_i$ $\in$ $\mathbb{Z}$ for
all $1$ $\leq$ $i$ $\leq$ $n$, $[u_1,$ $\ldots,$ $u_n]$ is an
increasing integer list by the initial hypothesis, and there
exists $1$ $\leq$ $i$ $\leq$ $n$ such that $\eta(X)$ $=$ $u_i$
$\in$ $\mathbb{Z}$, according to Table $1$ we can deduce
$domain^{\cdom}$ $\eta(X)$ $[u_1,$ $\ldots,$ $u_n]$ $\to$ $true$.
Moreover, since $\eta(X)$ $\in$ $\mathbb{Z}$, trivially
$indomain^{\cdom}$ $\eta(X)$ $\to$ $\top$ according again to Table
$1$. Then, $indomain^{\cdom}$ $\eta(X)$ $\to$ $\top$,
$domain^{\cdom}$ $\eta(X)$ $[u_1,$ $\ldots,$ $u_n]$ $\to$ $true$,
and we can conclude that $\eta$ $\in$ $Sol_{\cdom}(labeling$
$[\ldots]$ $[X],$ $X$ $\in$ $[u_{1},$ $\ldots,$ $u_{n}])$.
Therefore, $\eta$ $\in$ $Sol_{\cdom}(labeling$ $[\ldots]$ $[X],$
$X$ $\in$ $[u_{1},$ $\ldots,$ $u_{n}],$ $S$
$\myboxi$ $\sigma)$.\\\\
The remaining conditions of the theorem for this rule trivially
hold because of the initial hypothesis $labeling$ $[\ldots]$ $[X],$
$X$ $\in$ $[u_{1},$ $\ldots,$ $u_{n}],$ $S$ $\myboxi$ $\sigma$
satisfies the requirements of Definition $2$, and because of the conditions
of the rule $X$ $\notin$ $\chi$. The last rule for $labeling$
follows a trivial reasoning because $Sol_{\cdom}(labeling$
$[\ldots]$ $[u])$ $=$ $Val(\cdom)$ if $u$ $\in$ $\mathbb{Z}$.
According to Table $1$, $indomain^{\cdom}$ $u\eta$ $\to$ $\top$
for all $\eta$ $\in$ $Val(\cdom)$. Therefore,
$Sol_{\cdom}(labeling$ $[\ldots]$ $[u],$ $S$ $\myboxi$ $\sigma)$
$=$ $Sol_{\cdom}(labeling$ $[\ldots]$ $[u])$ $\cap$
$Sol_{\cdom}(S$ $\myboxi$ $\sigma)$ $=$ $Val(\cdom)$ $\cap$
$Sol_{\cdom}(S$ $\myboxi$ $\sigma)$ $=$ $Sol_{\cdom}(S$ $\myboxi$
$\sigma)$. The remaining conditions of the theorem are also trivial
by
initial hypothesis.\\\\
{\bf Failure Rules}\\\\
Finally, we suppose any arbitrary failure rule such that $S$
$\myboxi$ $\sigma$ $\red_{\chi}$ $fail$ and we prove that
$Sol_{\cdom}(S$ $\myboxi$ $\sigma)$ $=$ $\emptyset$. First, we
note that any failure rule must have the following syntactic form:
$S_1,$ $S_2$ $\myboxi$ $\sigma$ $\red_{\chi}$ $fail$ with
conditions such that $Sol_{\cdom}(S_1)$ $=$ $\emptyset$. For
example, consider the failure rule associated to Table $4$: $u$
$\leq$ $u',$ $S$ $\myboxi$ $\sigma$ $\red_{\chi}$ $fail$ with $u,$
$u'$ $\in$ $\mathbb{Z}$ and $u$ $>^{\mathbb{Z}}$ $u'$. Clearly,
$Sol_{\cdom}(u$ $\leq$ $u')$ $=$ $\emptyset$. In this situation,
$Sol_{\cdom}(S_1,$ $S_2$ $\myboxi$ $\sigma)$ $=$
$Sol_{\cdom}(S_1)$ $\cap$ $Sol_{\cdom}(S_2$ $\myboxi$ $\sigma)$
$=$ $\emptyset$ $\cap$ $Sol_{\cdom}(S_2$ $\myboxi$ $\sigma)$ $=$
$\emptyset$. \hfill $\blacklozenge$

\end{document}